\documentclass[floatfix,twocolumn,showpacs,preprintnumbers,amsmath,amssymb,prb,superscriptaddress,longbibliography]{revtex4-1}

\usepackage{graphicx}
\usepackage{dcolumn}
\usepackage{bm}
\usepackage{float} 
\usepackage{lipsum}
\usepackage{enumitem,amssymb} 
\usepackage{color}
\usepackage[usenames,dvipsnames,svgnames,table]{xcolor}
\usepackage{hyperref}

\usepackage{natbib}

\hypersetup{
    colorlinks = true,
    citecolor = blue
}
\newlist{todolist}{itemize}{2}
\setlist[todolist]{label=$\square$}

\usepackage{siunitx}

\usepackage{pifont}

\usepackage{leftidx}
\usepackage{tabulary}
\usepackage{tabularx,booktabs}
\usepackage[utf8]{inputenc}
\usepackage[T1]{fontenc}
\usepackage{csquotes} 

\newcommand{\sss}{\scriptscriptstyle}
\newcommand{\br}{\bm{r}}
\newcommand{\bR}{\bm{R}}
\newcommand{\bq}{\bm{q}}
\newcommand{\bk}{\bm{k}}

\newcommand{\s}{_\mathrm{{\scriptscriptstyle S}}}
\newcommand{\h}{_\mathrm{{\scriptscriptstyle H}}}
\newcommand{\xc}{_\mathrm{{\scriptscriptstyle XC}}}

\newcommand{\ofqw}{{(q,\omega)}}
\newcommand{\cg}{{\cal G}}

\begin{document}


\title{First-principles Modeling of Plasmons in Aluminum under Ambient and Extreme Conditions}

\author{Kushal Ramakrishna}
\email{k.ramakrishna@hzdr.de}

\affiliation{Helmholtz-Zentrum Dresden-Rossendorf (HZDR), D-01328 Dresden, Germany}
\affiliation{Technische  Universit\"at  Dresden,  D-01062  Dresden,  Germany}
\affiliation{Center for Advanced Systems Understanding (CASUS), D-02826 G\"orlitz, Germany}

\author{Attila Cangi}

\affiliation{Center for Advanced Systems Understanding (CASUS), D-02826 G\"orlitz, Germany}
\author{Tobias Dornheim} 

\affiliation{Center for Advanced Systems Understanding (CASUS), D-02826 G\"orlitz, Germany}

\author{Andrew Baczewski}
\affiliation{Center for Computing Research, Sandia National Laboratories, Albuquerque NM 87185 USA}

\author{Jan Vorberger}
\affiliation{Helmholtz-Zentrum Dresden-Rossendorf (HZDR), D-01328 Dresden, Germany}

\date{\today}
             
\begin{abstract} 
The theoretical understanding of plasmon behavior is crucial for an accurate interpretation of inelastic scattering diagnostics in many experiments. We highlight the utility of linear-response time-dependent density functional theory (LR-TDDFT) as a first-principles framework for consistently modeling plasmon properties. We provide a comprehensive analysis of plasmons in aluminum from ambient to warm dense matter conditions and assess typical properties such as the dynamical structure factor, the plasmon dispersion, and the plasmon lifetime. We compare our results with scattering measurements and with other TDDFT results as well as models such as the random phase approximation, the Mermin approach, and the dielectric function obtained using static local field corrections of the uniform electron gas parametrized from path integral Monte Carlo simulations. We conclude that results for the plasmon dispersion and lifetime are inconsistent between experiment and theories and that the common practice of extracting and studying plasmon dispersion relations is an insufficient procedure to capture the complicated physics contained in the dynamic structure factor in its full breadth.
\end{abstract}    
 
\maketitle

\section{Introduction} 
A consistent framework for modeling the properties of plasmons~\cite{PhysRev.82.625} in matter in the range from ambient to warm dense conditions is of utmost importance for both enhancing our fundamental understanding of extreme states of matter and supporting the diagnostics of scattering experiments~\cite{HED_frontiers}. As we will show, the properties of plasmons seem well understood under ambient conditions until one realizes that the data on plasmon dispersions and lifetimes are actually sparse and rarely consistent between different experiments and theories. Capturing plasmon dispersion and lifetimes in experiment and theory becomes even more challenging under warm dense conditions~\cite{wdm_book}. 

Warm dense matter (WDM) is highly energetic and exhibits characteristics of solids, liquids, and plasmas simultaneously~\cite{fortov_review,review,wdm_book,bonitz2019ab}.  
Understanding WDM is essential for enhancing our knowledge about astrophysical objects, such as the physics in Earth's core~\cite{AG98}, the formation processes of both planets in our solar system~\cite{Militzer_2008,militzer1,manuel, NBHR12,NHFR13, LHR09,LHR11} and of exoplanets\cite{NFKR11,KKNF12}, in brown dwarfs~\cite{saumon1,becker}, and stellar interiors~\cite{Daligault_2009}. 
From a technological point of view, warm dense conditions occur in the heating process of inertial confinement fusion capsules on their path towards ignition~\cite{AM04,hu_ICF} and in the walls of high-power magnetic fusion devices~\cite{SASAKI2015}.

Since both thermal and quantum effects have to be taken into account~\cite{KD09}, the quality of well-established methods of plasma physics or of condensed-matter physics might be insufficient. Hence, an understanding of plasmons under warm dense conditions relies on innovative theoretical tools and a close cooperation with experiments.

Experimental measurements of plasmons are carried out using several techniques such as X-ray Thomson scattering (XRTS)~\cite{Glenzer_plasmons}, electron energy-loss spectroscopy (EELS)~\cite{egerton2008electron}, and inelastic X-ray scattering (IXS) at synchrotrons~\cite{Mao_2001}. High pressures are induced using diamond anvil cells or (laser generated) shocks. 
Among these, the combination of high power optical lasers to generate WDM states and X-ray free electron lasers to diagnose them via XRTS is particularly useful~\cite{falk_2018}, especially for studying astrophysical phenomena in the laboratory.
Such an experimental setup is capable of achieving pressures on the order of a few Mbars and temperatures up to a few eVs. These experiments are nowadays performed at large-scale experimental facilities, such as SLAC~\cite{LCLS_2016} and the European XFEL~\cite{Tschentscher_2017}. Matter at even higher pressures and temperatures is investigated with highly energetic lasers at the NIF~\cite{Moses_NIF}.

The measured XRTS signal~\cite{Kraus2017,Knudson1455,Glenzer_plasmons,ernstorfer2,Fletcher2014,Frydrych2020} is directly linked to the numerical modeling of plasmons via the dynamic structure factor (DSF)~\cite{RevModPhys.81.1625}. The DSF is used as an important diagnostics for WDM~\cite{RevModPhys.81.1625}, because systems parameters like the density and temperature are inferred from it.

The DSF and, thus, plasmons have been modelled theoretically with a number of different techniques. The basic understanding of dispersion and damping of plasmons (in a gas of electrons) stems from the random phase approximation (RPA)~\cite{Kremp:1339576,thiele2008plasmon}. It was then realized that the influence of electron-ion correlations on the plasmon damping is essential. This led to the Mermin dielectric function featuring generalized dynamic collision frequencies of screened Born or T-matrix type~\cite{mermin1970lindhard,Reinholz_2000}. Likewise, electron-electron correlations are taken into account on a formally exact level by local field corrections (LFC). Accurate expressions for LFCs have recently been parametrized from quantum Monte Carlo (QMC) data\cite{doi:10.1063/1.5123013,dornheim2020effective,Dornheim_PRB_2020,Dornheim_PPCF_2020}.  Furthermore, combining electron-electron correlations with electron-ion correlations in dielectric theories was enabled in terms of the LFC-Mermin approach~\cite{Fortmann_2010}. As the collision frequencies of Born or T-matrix type were found to be insufficient for WDM, it was suggested to compute them from the Kubo-Greenwood approach~\cite{kubo1957statistical,greenwood1958boltzmann} and density functional molecular dynamics (DFT-MD) instead~\cite{PhysRevLett.118.225001,Ramakrishna_2019}.  

The logical extension of these prior plasma physics approaches is to simulate the system directly from first principles using techniques formerly used solely in solid-state physics. Dynamic properties can be obtained using linear response time-dependent density functional theory (LR-TDDFT). Here, Kohn-Sham (KS) orbitals incorporating the band structure are used instead of the free particle states as in the traditional plasma physics methods mentioned earlier. Then, similar to before, an approximation in terms of the response function with or without an exchange-correlation (XC) kernel can be established. The difficulty lies in finding reliable, accurate, temperature- and frequency-dependent XC functionals and XC kernels ~\cite{doi:10.1146/annurev.physchem.55.091602.094449,PhysRevB.84.075109,Ramakrishna_2019,bonitz2019ab,PhysRevB.101.195129}.

Real-time TDDFT is a method to compute the quantum dynamics governed by a time-dependent Hamiltonian.  It is distinct from LR-TDDFT and it enables one to go beyond the linear-response regime. However, it is worth noting that both should give the same results in the limit of the weak perturbations that we consider in this work~\cite{PhysRevLett.116.115004}.

In this paper, we present extensive LR-TDDFT results of the DSF, the dispersion, and lifetime of plasmons in aluminum under ambient and warm dense matter conditions. We use different kernels and the latest (temperature-dependent) LFC obtained from fermionic path integral Monte Carlo simulations (PIMC)~\cite{dornheim2019static}.
We compare these new results to a variety of theoretical and experimental data. We find insufficient agreement of our results with published data and many inconsistencies in the reported results. This means that the currently used XC functionals and kernels are not accurate enough to resolve remaining discrepancies between different experimental results.

Throughout this paper we work in atomic units, where $\hbar = m_e = e^2 = 1$, such that energies are expressed in Hartrees and length in Bohr radii.

\section{Methods}
Approximate solutions to the many-particle Hamiltonian can be found using methods like many-body perturbation theory, DFT, and TDDFT. Electronic transport properties such as the dynamic response function, the dielectric function, and the DSF are computed based on these methods. We first introduce the general formalism and our notation, then we describe the relevant methods in more detail.
\subsection{The Coupled Electron-Ion Problem}
Within the scope of non-relativistic quantum mechanics, the physics of coupled electrons and ions is governed by the many-particle Hamiltonian
\begin{align}
\hat{H} = \hat{H}^{\sss i} + \hat{H}^{\sss e} + \hat{W}^{\sss ei}\,,
\end{align}
with $\hat{H}^{\sss i} = \hat{T}^{\sss i} + \hat{W}^{\sss ii}$ denoting the kinetic energy and interaction of the ions, $\hat{H}^{\sss e} = \hat{T}^{\sss e} + \hat{W}^{\sss ee}$ the kinetic energy and interaction of the electrons, and $\hat{W}^{\sss ei}$ the interaction between the electrons and ions. 

Furthermore, working within the Born-Oppenheimer approximation~\cite{AbMaGr2012} reduces the solution of the coupled electron-ion problem to solving a Schrödinger equation for the electrons 

\begin{align}
\label{eq:electrons.SE}
& \hat{H}^{BO}(\br_1,...,\br_{N_e}; \bR_1,...,\bR_{N_i})\ \Psi_j(\br_1,...,\br_{N_e})\nonumber\\ 
& = E^{BO}_j(\bR_1,...,\bR_{N_i})\ \Psi_j(\br_1,...,\br_{N_e}),
\end{align}
which depends parametrically on the coordinates of the underlying ionic structure through the potential energy surface $E^{BO}_j(\bR_1,...,\bR_{N_i})$. Here, the Born-Oppenheimer Hamiltonian is given by $\hat{H}^{BO} = \hat{T}^{e} + \hat{W}^{ee} + \hat{W}^{ei} + \hat{W}^{ii}$, where the $N_e$ electrons have coordinates $\br_j$, while the $N_i$ ions have mass $M_I$, charge $Z_I$, and coordinates $\bR_I$. 
\subsection{Dielectric Response of the Born-Oppenheimer Hamiltonian}
The \emph{linear} response $n_{\sss ind}(\bq, \omega)$ of the electronic system defined by the Born-Oppenheimer Hamiltonian in Eq.~(\ref{eq:electrons.SE}) to an external, time-dependent perturbation $\delta v$ is given in Fourier space by
\begin{equation}\label{eq:n1.chi.ex.ft}
n_{\sss ind}(\bq, \omega) = \chi(\bq, \omega) \delta v(\bq, \omega)\,,
\end{equation}
where the proportionality factor corresponds to the density-density response function $\chi(\bq, \omega)$.
The dielectric function $\epsilon(\bq, \omega)$ is expressed in terms of the density-density response function as
\begin{equation}\label{eq:n1.chi.eps}
\frac{1}{\epsilon(\bq, \omega)} = 1 + \frac{4\pi}{\bq^2} \chi(\bq, \omega)\ .
\end{equation}
Furthermore, the fluctuation-dissipation theorem~\cite{So2010} connects the DSF to the density-density response function 
\begin{equation}\label{eq:n1.s.chi}
S(\bq,\omega)=-\frac{1}{\pi n_{e}\left(1-e^{-\omega/(k_{B} T_{e})}\right)}\mathrm{Im}\left[\chi(\bq,\omega)\right],
\end{equation}
and, hence, to the dielectric function~\cite{hamann2020dynamic}
\begin{equation}
S(\bq,\omega) = -\frac{\bq^{2}}{4\pi^{2}n_{e}\left(1-e^{-\omega/(k_{B} T)}\right)}
\mathrm{Im}\left[\epsilon^{-1}(\bq,\omega)\right], 
\end{equation}   
where $n_{e}$ is the free electron density and $T$ the temperature. Note that throughout the paper, the theoretical methods we use assume equilibrated temperature for ions and electrons. Hence, we use the term \emph{temperature} to refer to both electronic and ionic temperature. Using the detailed balance relation for the DSF $ S(-\bq,-\omega) = S(\bq,\omega)  e^{ - \beta \hbar \omega } $, diagnostics of parameters in experiments such as the temperature, the equation of state, the ionization potential, and the density are inferred~\cite{RevModPhys.81.1625}. 
Traditionally, the DSF for WDM and high energy density matter is modelled using plasma physics based theories and various approximations. Recently  \emph{first-principles} methods based on TDDFT have been successful in modeling XRTS spectra~\cite{PhysRevLett.116.115004,PhysRevLett.120.205002, Ramakrishna_2019,Frydrych2020}.

The imaginary part of the inverse dielectric function is the spectral function of collective excitations in the system. Particularly important among these are plasmons which appear in the DSF as sharp peaks near the plasma frequency. They can be mathematically characterized as zeros of the complex dielectric function, emerge in the parameter range for which collective effects play a role in the response of the system, and are the dominant mechanism for very small wavenumbers~\cite{Kremp:1339576,Bonitz:2016,hamann2020ab}. 

\subsection{Plasmon Properties from Dielectric Models and Time-dependent Density Functional Theory}

\subsubsection{Random-Phase Approximation}
The RPA has been applied widely in condensed matter and plasma physics to describe collective excitations. The RPA is the simplest approximation capable of describing collective properties of a system of weakly interacting electrons (jellium model) immersed in a uniform positively charged background~\cite{PhysRev.82.625,pines1952collective,bohm1953collective}.

The form of the RPA retarded dielectric function based on free electron states as used here is discussed in Refs.~\cite{Kremp:1339576,QSOCPS,BaymQSM}. 
The imaginary and the real parts of the dielectric function are given by
 
\begin{multline}
\Im [\varepsilon(q, \omega)]  =  (2s_{e}^{z}+1) \dfrac{4 \pi m^2 e^{2} k_BT}{ \hbar q^{3} } \\ 
\times \ln\left\{
\frac{1+\exp\left\{\beta\left(-\frac{E^-}{2m}+\mu\right)\right\}}
{
1+\exp\left\{\beta\left(-\frac{E^+}{2m}+\mu\right)\right\}
}
\right\}, 
\end{multline}  
and
\begin{multline}
\Re [\varepsilon(q, \omega)]  =  1 - (2s_{e}^{z}+1) \dfrac{16 \pi^{2}  m e^{2} \hbar^{2} }{ q^{2} }\,  {\cal P}\!\!\! \int\displaylimits_{ -\infty }^{ +\infty } \dfrac{ d p }{ (2 \pi \hbar)^{3} }\, p f(p)
 \\
\times  \dfrac{1}{2q}  \left[ \log \left( \dfrac{ p a_{b} }{ \hbar } - \dfrac{ q a_{b} }{ 2 \hbar } - \dfrac{ m \omega  a_{b} }{ q }  \right )  
  \right. \\ 
+ \left. \log \left ( \dfrac{ p a_{b} }{ \hbar } - \dfrac{ q a_{b} }{ 2 \hbar } + \dfrac{ m \omega  a_{b} }{ q }  \right )   \right]  , 
\end{multline}

where $m$ is the electron mass, $s_{e}^{z}$ accounts for the spin of the electrons, and $E^{\pm}=(\pm q/2-m\hbar\omega/q)^2$.
The plasmon dispersion relation within the RPA and for small wavenumbers is given by~\cite{thiele2008plasmon,hamann2020ab} \begin{equation}
\omega^{2}(q) = \omega^{2}_{pl}  \Bigg  [ 1 + \frac{ \langle p^{2} \rangle  }{ m^{2} } \frac{ q^{2} }{\omega^{2}(0)} + \frac{ \langle p^{4} \rangle  }{ m^{4} } \frac{ q^{4} }{\omega^{4}(0)}  +\ldots \bigg ] ,
\label{plasmon_dispersion}
\end{equation}
where $\omega_{pl}$ is the plasma frequency with the moments $\langle p^{i} \rangle$ evaluated using the Fermi integral~\cite{galassi2002gnu}. The plasmon dispersion can be obtained experimentally by fitting the data as obtained from EELS and IXS experiments to this expression~\cite{Krane_1978,RevModPhys.81.1625}. 

\subsubsection{Extended Mermin Dielectric Function}
The RPA is not sufficient to account for strong correlations and the jellium model is not sufficient to account for bound states. Introducing a dynamic damping or relaxation term $\nu(\omega)$, while maintaining density conservation, leads to the Mermin approach (MA)~\cite{mermin1970lindhard,Reinholz_2000} 
\begin{equation}
 \epsilon^{MA}(q,\omega) = 1 + \dfrac{ \left[1 + i \nu(\omega)/\omega\right] \left[ \epsilon\left(q,\omega + i \nu(\omega)\right) - 1\right] }{ 1 + \left[i \dfrac{\nu(\omega)}{\omega}\right] \dfrac{  \epsilon\left(q,\omega + i \nu(\omega)\right) - 1    }{  \epsilon(q,\omega \rightarrow 0) - 1  }   } , 
 \label{eq:xMA}
\end{equation}
which takes into account electron-ion collisions.
The dielectric function $\epsilon\left(q,\omega + i \nu(\omega)\right)$ may be taken in RPA using the dynamic collision frequency. The latter is obtained in screened Born or T-matrix approximation~\cite{Reinholz_2000}, or is computed using the Kubo-Greenwood approach based on KS orbitals and eigenvalues~\cite{PhysRevLett.118.225001,Ramakrishna_2019}. 

\subsubsection{Local Field Corrections}

Local field corrections (LFC) in isotropic systems like fluids or plasmas are defined such that the full response function can be obtained from a convolution of the free density response function $\chi^{0}\ofqw$ and the LFC $G(q,\omega)$
\begin{eqnarray}\label{eq:define_LFC}
\chi({q},\omega) = \frac{ \chi^0({q},\omega) }{ 1 - V(q)\big[1-G({q},\omega)\big]\chi^0({q},\omega)}\ .
\end{eqnarray}
The corresponding equation for the dielectric function reads
\begin{equation}
 \epsilon(q,\omega) = 1 - \dfrac{ 1 - \epsilon^{RPA}(q,\omega)  }{ 1 + G(q,\omega)\left[1 - \epsilon^{RPA}(q,\omega)\right]  }\ .
\end{equation}
Such an expression can be used to incorporate strong electron-electron correlations into the Mermin approach Eq.~(\ref{eq:xMA}), which leads to a dielectric function of extended Mermin type~\cite{Fortmann_2010}. 
 
Since the electronic LFC of a realistic system like warm dense aluminum intrinsically depends on the ionic component as well, the full problem typically cannot be solved. Therefore, one often substitutes the correct $G(q,\omega)$ of the full system by the LFC of an uniform electron gas at the same density and temperature. 

Often QMC data for the LFC, and representations thereof, are restricted to the static limit $\omega=0$. While such a static approximation would, in principle, constitute an uncontrolled approximation, it has recently been shown~\cite{PhysRevLett.121.255001,PhysRevB.99.235122,dornheim2020finitesize} that the frequency dependence of $G(q,\omega)$ has a negligible impact for $r_s\lesssim4$, which is the case for the conditions considered in this work. 

\begin{figure}[t]          
\centering       
\includegraphics[width=1.0\columnwidth]{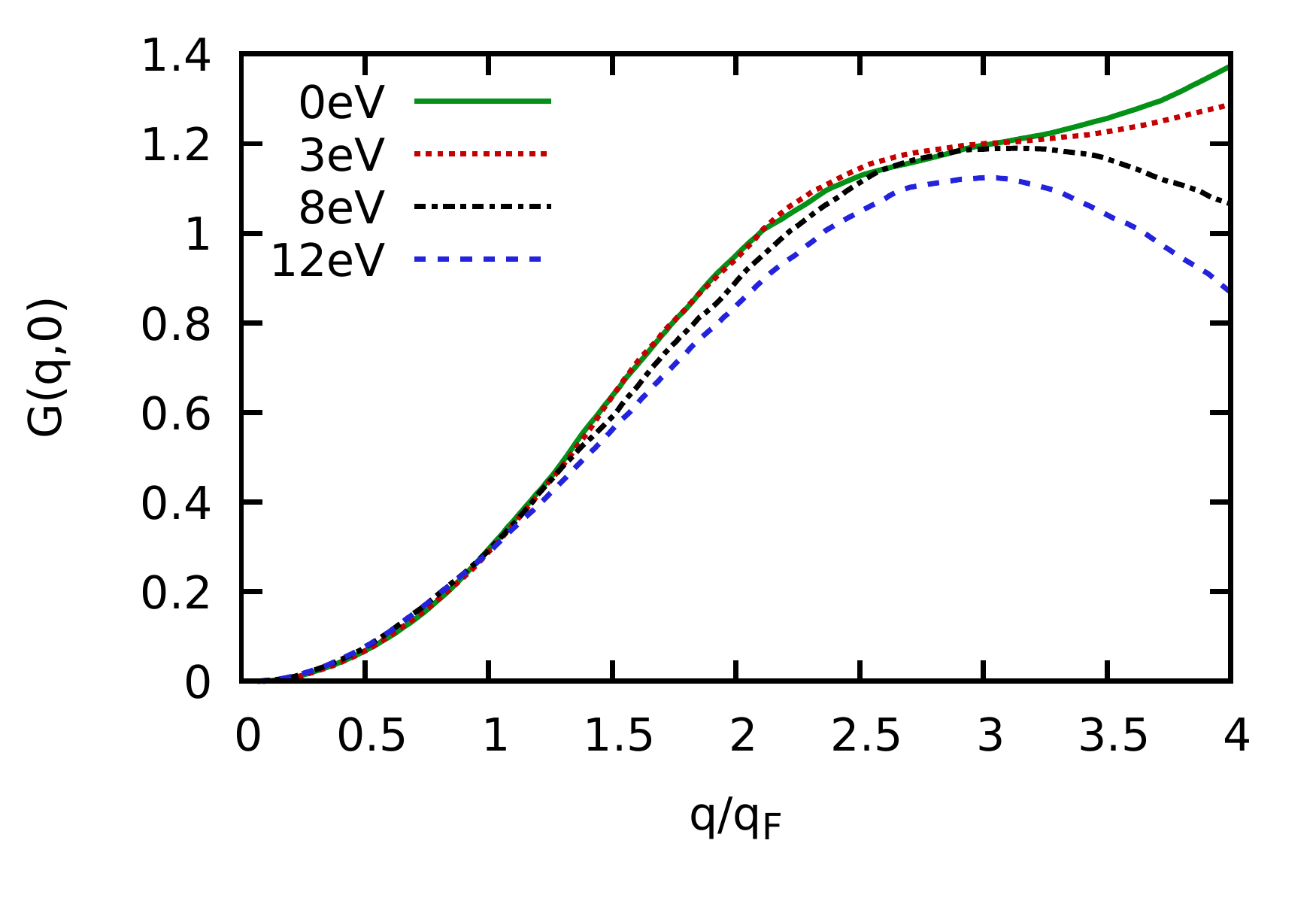}
\caption{\raggedright Static LFC of aluminum ($r_s=2.07$) at four different temperatures. The data have been obtained from the machine-learning representation from Ref.~\cite{doi:10.1063/1.5123013}.  }    
\label{Aluminum_LFC}
\end{figure}    

The first accurate data for the static LFC $G(q,0)$ of the UEG have been obtained by Moroni \textit{et al.}~\cite{PhysRevLett.69.1837,PhysRevLett.75.689} on the basis of ground-state QMC simulations. These data have subsequently been parametrized by Corradini \textit{et al.}~\cite{PhysRevB.57.14569} (CDOP), and have been widely used to include electronic correlation effects in many-body theory. Unfortunately, their parametrization is limited to the zero temperature limit, which is often not sufficient for realistic WDM applications~\cite{PhysRevE.93.063207}. This problem has been overcome only recently by Dornheim \textit{et al.}~\cite{doi:10.1063/1.5123013,dornheim2020effective,Dornheim_PRB_2020,Dornheim_PPCF_2020}, who presented a machine-learning representation of the static LFC (hereafter denoted as T-LFC) with respect to $r_s$, $\Theta=K_BT/E_F$, and $q$ on the PIMC data at finite temperature, covering the entire WDM regime, where $E_F$ denotes the Fermi energy and $K_B$ the Boltzmann constant.

The T-LFC is illustrated in Fig.~\ref{Aluminum_LFC} at the density of aluminum ($r_s=2.07$) for four different temperatures. For $T=0$ (solid green), the ML-representation reproduces the ground-state parametrization of CDOP. At $T=3$ eV (dashed red, $\Theta\approx0.26$), the effect of the temperature on $G$ is small and only starts to manifest at large wavenumbers. Upon further increasing the temperature to $T=8$ eV (dash-dotted black, $\Theta\approx0.69$) significant deviations from the ground-state result are apparent, which are particularly pronounced for large $q$ where the tail of the LFC (i.e., the asymptotic behaviour of $G(q)$ for large wave numbers) shows a negative slope and eventually becomes negative. This is related to a lowering of the kinetic energy due to XC effects at these conditions, see Ref.~\cite{doi:10.1063/1.5123013} for an extensive discussion. Finally, the largest deviations are observed at $T=12$~eV (dashed blue, $\Theta\approx1.03$), where $G$ is systematically lower than at $T=0$ for wavenumbers higher than the Fermi wavenumber. 
Therefore, we expect temperature effects of the LFC to be observable in our simulation results for temperatures $T\gtrsim 8$~eV and wavenumbers $q>q_F$.

\subsubsection{Density Functional Theory Coupled to Molecular Dynamics}
In the framework of KS-DFT~\cite{KoSh1965}, a solution to Eq.~(\ref{eq:electrons.SE}) is found in a computationally feasible manner by introducing a fictitious system of non-interacting electrons that yields the same electronic density as obtained from directly solving Eq.~(\ref{eq:electrons.SE}). This is achieved by solving a set of KS equations
\begin{equation}
\label{eq:ks.dft}
\left[ -\frac{1}{2} \nabla_k^2 + V\s(\br)\right] \phi_k(\br) = \epsilon_k \phi_k(\br) , 
\end{equation}
for the KS orbitals $\phi_k$ from which the electronic density is constructed according to 
$n(\br) = \sum_k^{N^{e}} \phi^*_k(\br)\,\phi_k(\br)$.

Note that the solutions of the KS equations have a parametric dependence on the underlying ionic configuration $\{ \bR_1,...,\bR_{N_i}\}$ via the KS potential
$v\s(\br; \bR_1,...,\bR_{N_i})  = \sum_{I=1}^{N_i} Z_I/|\br_k-\bR_I| 
+ v\h[n](\br) + v\xc[n](\br)$, 
where 
$v\h[n](\br) = \int d\br' n(\br)/|\br-\br'|$ 
denotes classical electrostatic interaction potential of a charge cloud (Hartree potential) and 
$v\xc[n](\br) = \delta E\xc[n]/\delta n(\br)$ 
the XC potential. While formally exact, in practice the XC energy $E\xc[n]$ is unknown and approximations need to be used~\cite{PrGrBu2015}. Furthermore, KS-DFT is generalized to finite temperature via Mermin's theorem~\cite{Mermin_1965}. We follow the common approximation, where the explicit temperature dependence of the XC energy is neglected and only the implicit temperature dependence in the electronic density is taken into account. To that end, the temperature-dependent density is computed from the KS orbitals as 
$n(\br) = \sum_k^{\infty} f_k(T) \phi^*_k(\br)\,\phi_k(\br)$, 
where $f_k(T) $ denotes the Fermi-Dirac distribution at temperature $T$.
 
\subsubsection{Time-dependent Density Functional Theory}
LR-TDDFT~\cite{GrKo1985} is a commonly used method to compute electronic response properties in a sufficiently accurate and computationally feasible manner. The formally exact, linear density-density response of the electronic system defined by the Born-Oppenheimer Hamiltonian given in Eq.~(\ref{eq:electrons.SE}) to an external, time-dependent perturbation $\delta v(\br, t)$ is given as  
\begin{equation}
\chi_{\cg\cg'}(\bq,\omega) = \frac{ \chi^{KS}_{\cg\cg'}(\bq,\omega) }{1 - \left[V(\bq)+f\xc(\bq,\omega)\right]\chi^{KS}_{\cg\cg'}(\bq,\omega)  }\,
\label{chi_eqn}
\end{equation} 
where $\chi^{KS}_{\cg\cg'}$ denotes the KS density-density response function~\cite{doi:10.1146/annurev.physchem.55.091602.094449,ullrich2014brief}
\begin{multline}
\chi^{KS}_{\cg\cg'}(\bq,\omega)=
-\frac{1}{V}\lim_{\eta\to0^+}
\sum_{nm;\bk}\left[f_{m;\bk+\bq}(T)-f_{n;\bk}(T)\right]
\\
\times
\frac{\langle \psi_{m;\bk+\bq}|e^{i(\bq+\cg)\br}|\psi_{n;\bk}\rangle
\langle\psi_{n;\bk}|e^{-i(\bq+\cg')\br'}|\psi_{m;\bk+\bq}\rangle}
{\omega-\epsilon_{m;\bk+\bq}+\epsilon_{n;\bk}+i\eta} ,
\label{chi0} 
\end{multline}
defined in terms of the KS orbitals, eigenvalues, and Fermi occupations given by $f(T)$. $\eta\to 0^+$ is the Lorentzian broadening. $V(\bq) = 4 \pi \delta(\cg-\cg')/(|(\cg+\bq)(\cg'+\bq)|)$ is the Coulomb potential with $\cg$, $\cg'$ being the reciprocal lattice vectors~\cite{Sharma2014}. The DSF in this work is given by the macroscopic response functions which are obtained using $\chi(\bq,\omega)_{\cg=0,\cg'=0}$.

Adler-Wiser local field {\em effects} for the ideal lattice conditions are thus included by default~\cite{Adler1962quantum,Wiser1963}. Higher order electron-electron correlations are represented by the XC kernel which is formally defined as 
\begin{equation}\label{eq:fxc}
f\xc(\bq,\omega) = {\chi^{KS}}^{-1}(\bq,\omega)-\chi^{-1}({\bf
q},\omega)-v(\bq)\ .
\end{equation}
It is related to the XC potential via
$f\xc(\bq,\omega) = \delta v\xc(\bq,\omega)/\delta n(\bq, \omega)$
and to the LFC of dielectric models via 
$f\xc(\bq,\omega) = -v(\bq) G(\bq, \omega)$,
where $v(\bq)=4\pi/\bq^2$.

Virtually all practical calculations in LR-TDDFT employ a static (i.e., frequency-independent) $f\xc$, usually using the adiabatic local density approximation (TDDFT-XC, ALDA). Further neglecting the XC kernel, i.e., $ f\xc \rightarrow 0 $, yields what is called RPA calculations within the LR-TDDFT framework (TDDFT-RPA). When we employ a LFC from quantum Monte Carlo calculations or other methods, we call the results TDDFT-LFC.

Then, plasmon properties such as the DSF are computed within the LR-TDDFT from the density-density response function through the fluctuation-dissipation theorem as before.

\subsection{Computational Workflow}
While we use the well-established ideal crystal structure for ambient conditions, we run DFT-MD simulations (usually using VASP~\cite{PhysRevB.47.558,PhysRevB.59.1758,KRESSE199615,PhysRevB.54.11169}) for warm dense or high pressure conditions in order to generate snapshots of ionic configurations. 
A number of these supercells of up to $N=32$ ions are then subject to a high-resolution DFT calculation. Based on the resulting KS orbitals and various choices of LFCs and XC kernels, the density response function and, hence, plasmon properties are computed. 
Further details on all the technical parameters, settings, and certain convergence test results can be found in Appendix~\ref{app_Dl}.  

\section{Results}     

In what follows, we analyze the aluminum DSF under a variety of thermodynamic conditions, comparing to prior experiments and LR-TDDFT calculations.
We first consider ambient conditions, allowing us to establish an understanding of the effect that various approximations have on the results of our calculations relative to a tightly constrained temperature and ionic configuration.
Here aluminum adopts its standard face-centered cubic (fcc) crystal structure for which the electronic structure is well described by band theory and the collective plasmonic excitations are described in terms of standard condensed matter terminology.
We then study the aluminum DSF under more extreme conditions under which this picture might break down.
It may be the case that the crystalline order is destroyed, in which case short-range electron-ion correlations provide the dominant contribution to the physics beyond a picture of aluminum as a uniform electron gas.
In this context we adopt language that is more typical in the warm dense matter literature in which correlations, collisions, or interactions between electrons and ions give rise to ``electron-ion correlations'' that impact the behavior of plasmons.

We are careful to note the distinction between TDDFT-RPA and RPA. Without the TDDFT prefix, RPA generically refers to response functions calculated using model dielectric functions. With the TDDFT prefix it refers to atomistic LR-TDDFT calculations in which $G(q)=f_{xc}=0$.

\begin{figure*}[htp] 
\centering   
\includegraphics[width=1.0\columnwidth]{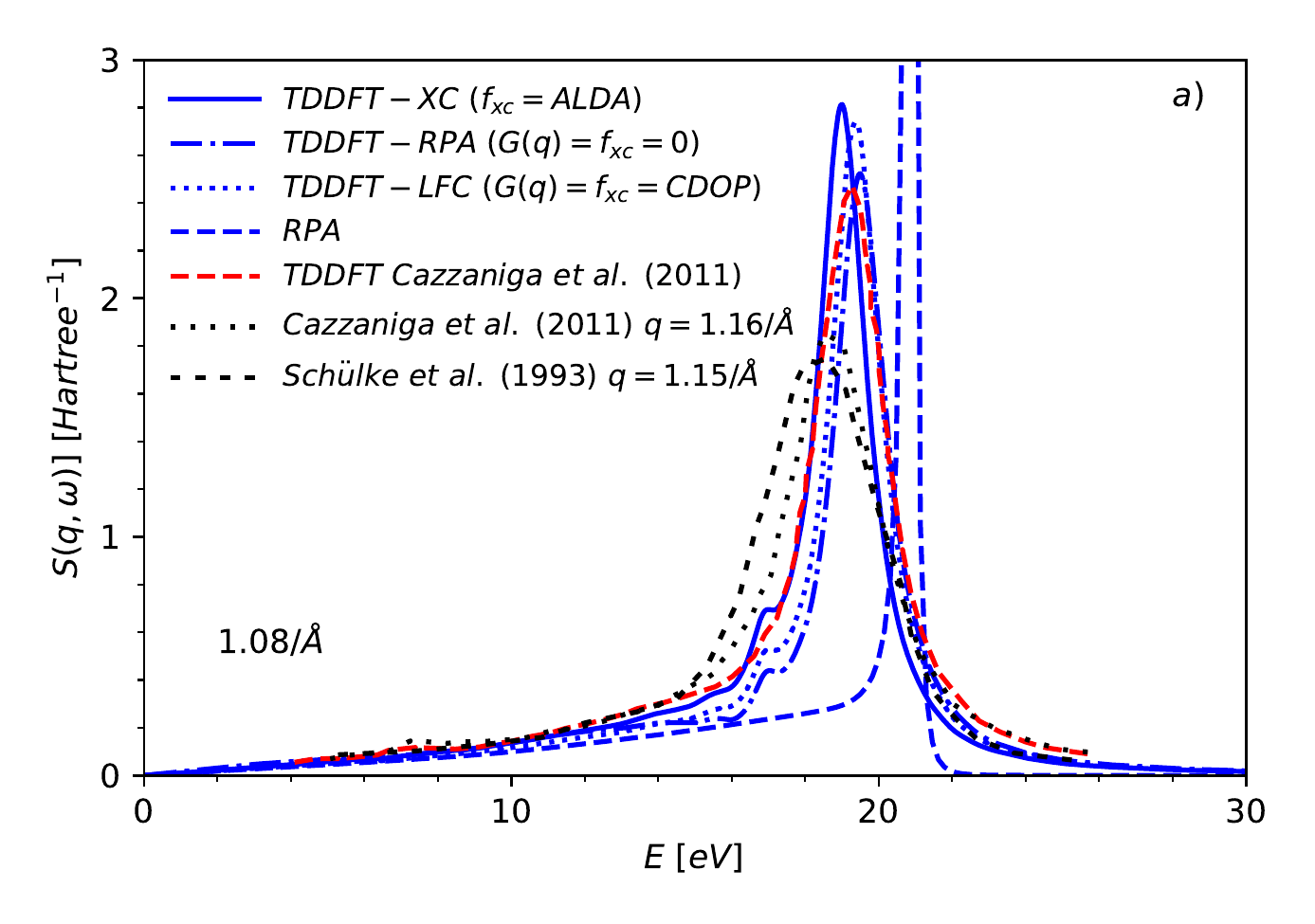} \includegraphics[width=1.0\columnwidth]{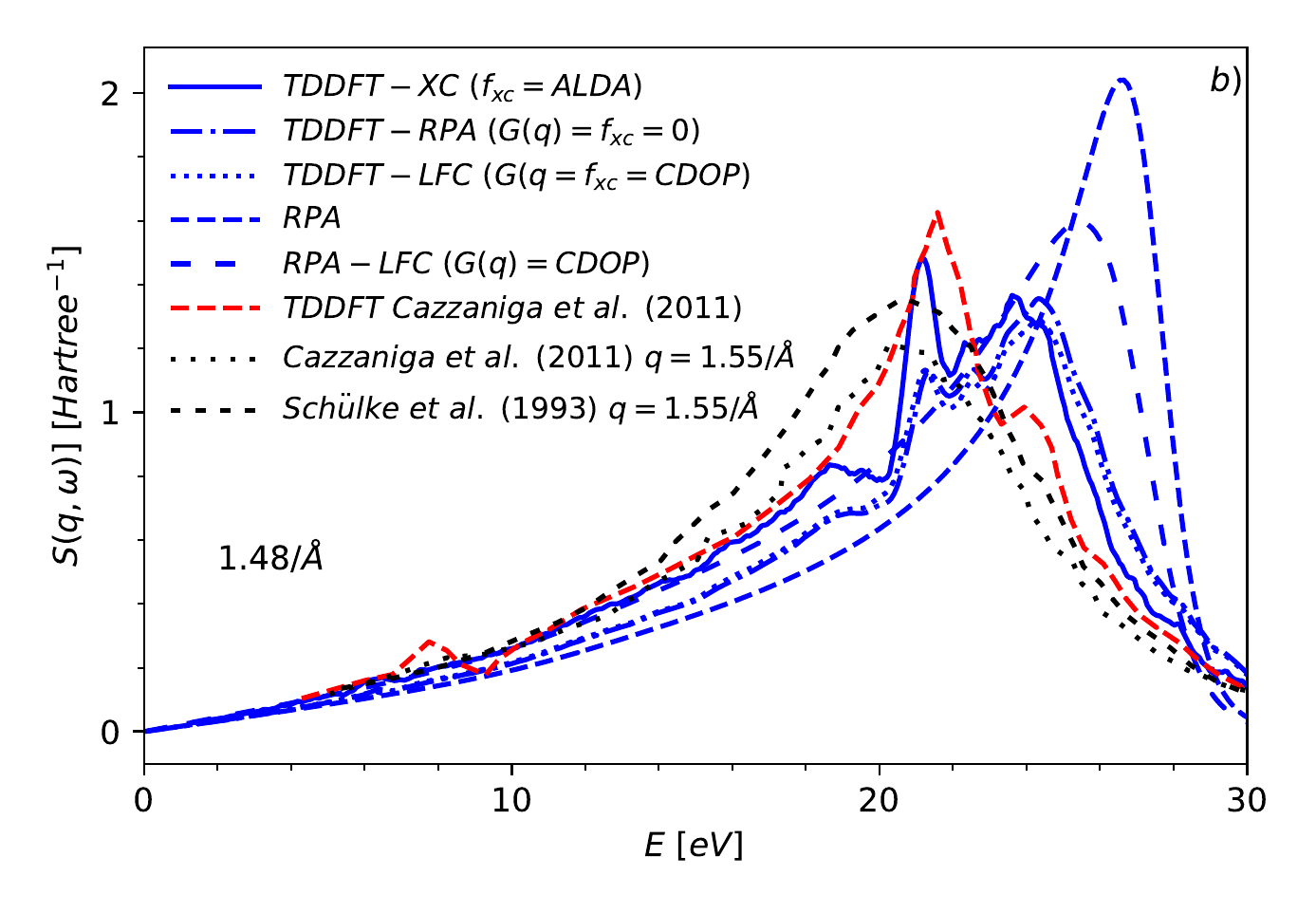}
\includegraphics[width=1.0\columnwidth]{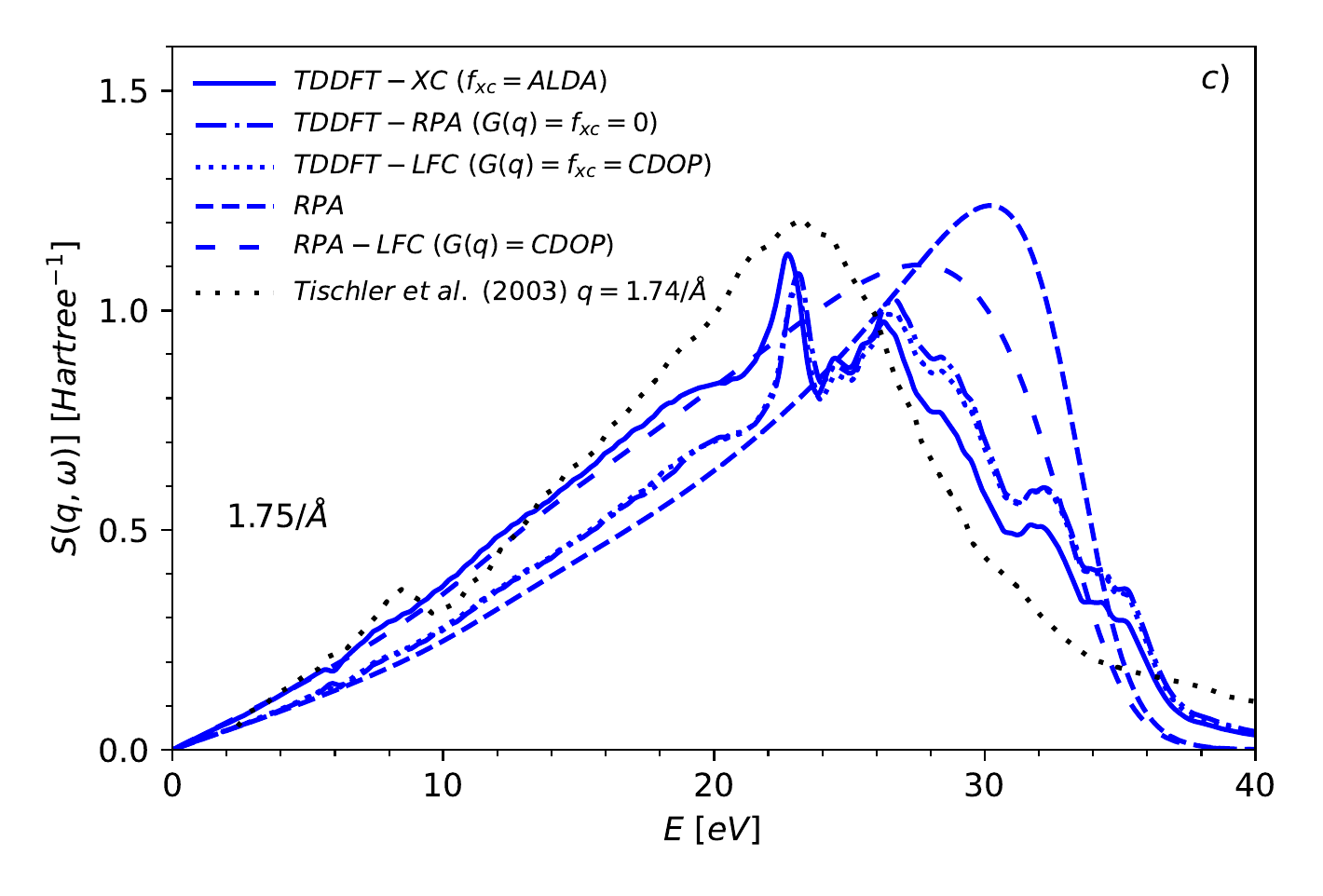}
\includegraphics[width=1.0\columnwidth]{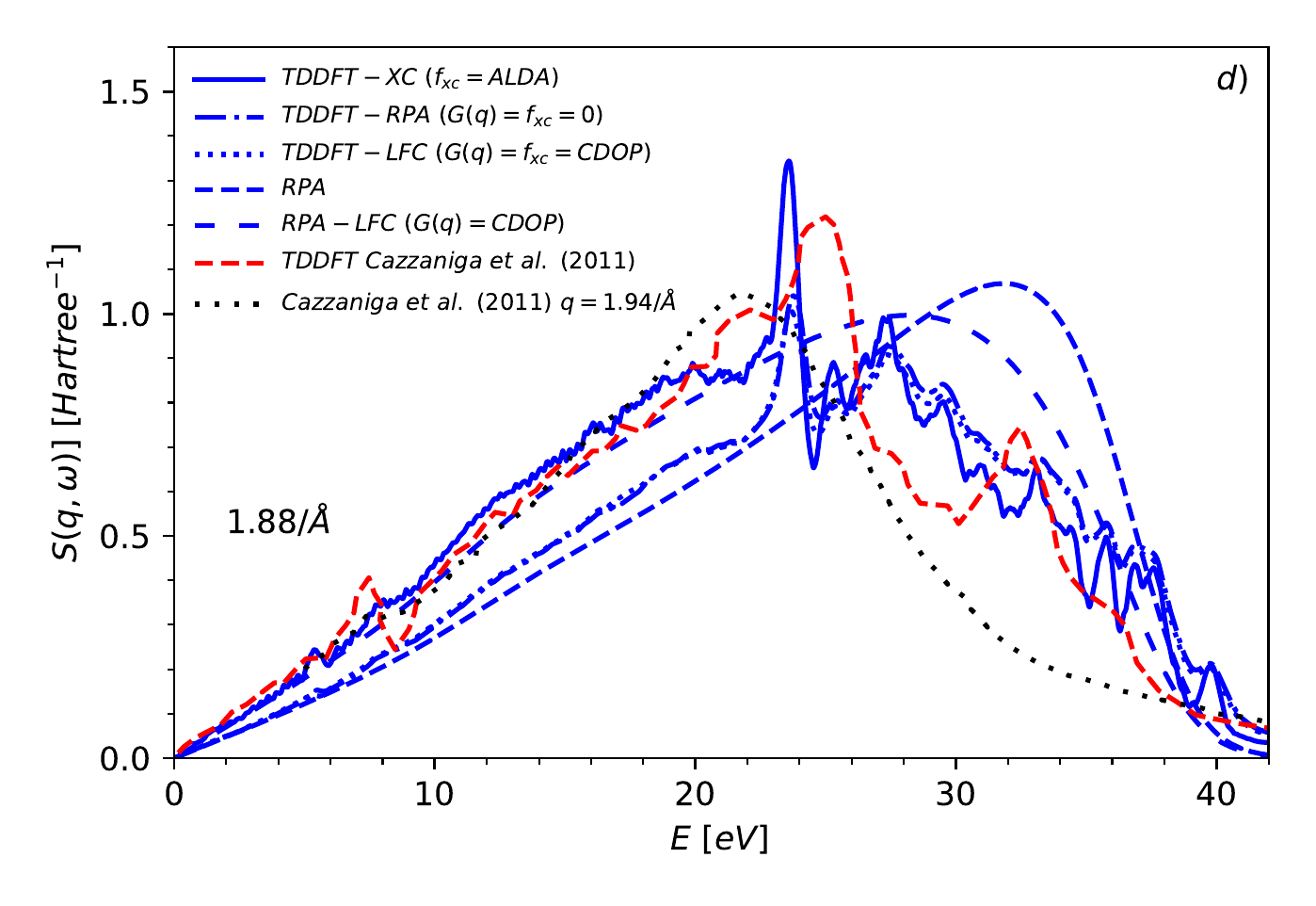} \caption{\raggedright DSF for aluminum in atomic units under ambient conditions at a) 1.08, b) 1.48, c) 1.75, and d) 1.88 (\si{\angstrom^{-1}}). Experimental data (in black) stems from Ref.~\onlinecite{PhysRevB.84.075109,PhysRevB.47.12426,test123}. Theoretical data (in red) stems from Ref.~\onlinecite{PhysRevB.84.075109}. TDDFT-RPA/TDDFT-XC/TDDFT-LFC/RPA/RPA-LFC results of this work are shown in blue. The scattering parameter $\alpha=\kappa/q$, with $\kappa$ being the inverse screening length, is unity for $q=2.04/$\si{\angstrom}, collective effects dominate for $\alpha\ll 1$~\cite{RevModPhys.81.1625,hamann2020ab}.} \label{Al_dsf_ambient_conditions}  
\end{figure*}        
  
\subsection{Ambient Conditions}
At standard temperature and pressure, solid aluminum adopts a fcc crystal structure with lattice constant 4.05 \AA~and density 2.7 g/cm$^3$. With a Seitz radius of 2.07 atomic units and a density of states resembling the free electron gas, one would expect the plasmon dispersion to be described well by the RPA~\cite{Ashcroft76}. 

\subsubsection{Dynamic Structure Factor}
The DSF at ambient conditions (T = 300 K) for a range of wavenumbers is shown in Fig.~\ref{Al_dsf_ambient_conditions}, where our calculations are compared to the nearest set of wavenumbers available in the literature~\cite{PhysRevB.84.075109,PhysRevB.47.12426,test123}.  

We start our discussion with panel a) at a wavenumber $1.08$\si{\angstrom^{-1}}
for which the system is clearly dominated by collective effects, and the sharp plasmon carries most of the spectral weight. 

The peak of the free electron RPA result is roughly $2$~eV above both experiments and all of the adiabatic TDDFT calculations.
That the TDDFT-RPA result gives a plasmon peak that is more consistent with these results suggests that deviations from a free electron model due to the lattice are a critical feature of the density response.
In fact, the TDDFT-XC and TDDFT-LFC results both agree well with TDDFT-RPA, suggesting that exchange-correlation effects and local field corrections do not matter as much as capturing deviations from the free electron model due to band structure effects at this value of $|\bq|$.
All three of these results stem from the same initial set of KS orbitals and thus capture the deviations from a free electron model in the same way.

Interestingly, TDDFT-XC calculations by Cazzaniga {\em et al.}~\cite{PhysRevB.84.075109} yielded an identical peak location, but a rather smaller peak height. This means that in the energy range of the plasmon, Cazzaniga's imaginary part of the response function is about $10$ percent larger than in our calculations. The main difference between our results and those of Cazzaniga {\em et al.} is that ours are obtained using an all-electron code and Cazzaniga {\em et al.} use a norm-conserving pseudopotential.

Even more different is the shape and location of the experimentally determined plasmon peaks for this wavenumber. The two different experimental results are in good agreement with each other~\cite{PhysRevB.84.075109,PhysRevB.47.12426}. However, they are another factor of $1.5$ smaller than Cazzaniga's and therefore only $60$ percent as high as our results. Its location is lower by another eV. The differences appear even more pronounced if one takes into account that the considered wavenumbers are not identical.

There are a few possible sources for the disagreement between experiment and theory. Cazzaniga {\em et al.} noted that the use of an adiabatic XC kernel might be one, and they report an improvement in the agreement when incorporating lifetimes from GW calculations~\cite{PhysRevB.84.075109}. However, even this procedure does not give good agreement between theory and experiment for all wavenumbers. A non-local and dynamic Kernel as suggested by Panholzer {\em et al.} might give improvements~\cite{PhysRevLett.120.166402}. However, there is also always the possibility that the KS orbitals (and therefore the XC functional) are not good enough when they are to be used to calculate the KS response function.

Increasing the wavenumber of the perturbation as shown in panels b) to d) leads to a broadening of the plasmon peak and finally to a mix of collective and single-particle effects which all contribute to the DSF. 
As was the case in panel a), the TDDFT peaks occur at lower energies than for the RPA. 
The influence of the LFC is best visible at large \emph{q} when compared to the electron gas RPA, with lower intensities at the peaks and a shift towards lower energies at small \emph{q}. In the TDDFT results, the difference between no LFC and different types of LFCs (ALDA or CDOP) is less distinct, still the same trend of redshift remains.
The effect of the LFC in aluminum has been determined experimentally by Larson \textit{et al.}~\cite{PhysRevLett.77.1346} for \emph{q} up to $4.37/ \si{\angstrom}$. They suggest a stronger impact at large wavenumbers as predicted by calculations with LFCs.

The overall shape of the spectra in Fig.~\ref{Al_dsf_ambient_conditions} c) to d) continues to differ from the experimental results. While the maximum intensity is now in better agreement with the theoretical results, the TDDFT curves start to show a double peak structure still absent in the experimental curves displayed here. The overall peak position in our results remains shifted to higher energies as compared to the experimental and Cazzaniga's theoretical results. The disagreement is even more worrying, when considering the higher number of bands, k-points and the number of explicitly treated electrons that are taken into account in our calculations as compared to the earlier published results.

Usually, a two peak structure, as it seems to emerge from TDDFT at the higher wavenumbers in Fig.~\ref{Al_dsf_ambient_conditions}, is associated with plasmon and double plasmon excitations. It is already accounted for by the non-interacting electron-hole bubble (and the band structure) and does not need higher order Coulomb correlations to appear. However, it seems that our TDDFT calculations overestimate the double plasmon excitations as in experiments they appear only at larger wavenumbers.

Inclusion of many-body effects, in the form of vertex corrections, might improve the agreement with the experimental measurements at large \emph{q}~\cite{fleszar1995band}. The inclusion of a nonlocal and dynamical XC kernel in TDDFT is further shown to improve the DSF in some metals and semiconductors including the double-plasmon excitation~\cite{PhysRevLett.120.166402,PhysRevLett.95.157401,huotari2008electron,petri1976anisotropy}. Sturm \textit{et al.}~\cite{PhysRevB.62.16474,pandey1974plasmons} demonstrated that at large \emph{q} and large frequencies dynamical correlations in $f_{xc}$ are more important than band structure effects in the description of the DSF. 

\subsubsection{Plasmon Dispersion} 

\begin{figure}[th]         
\centering           
\includegraphics[width=1.0\columnwidth]{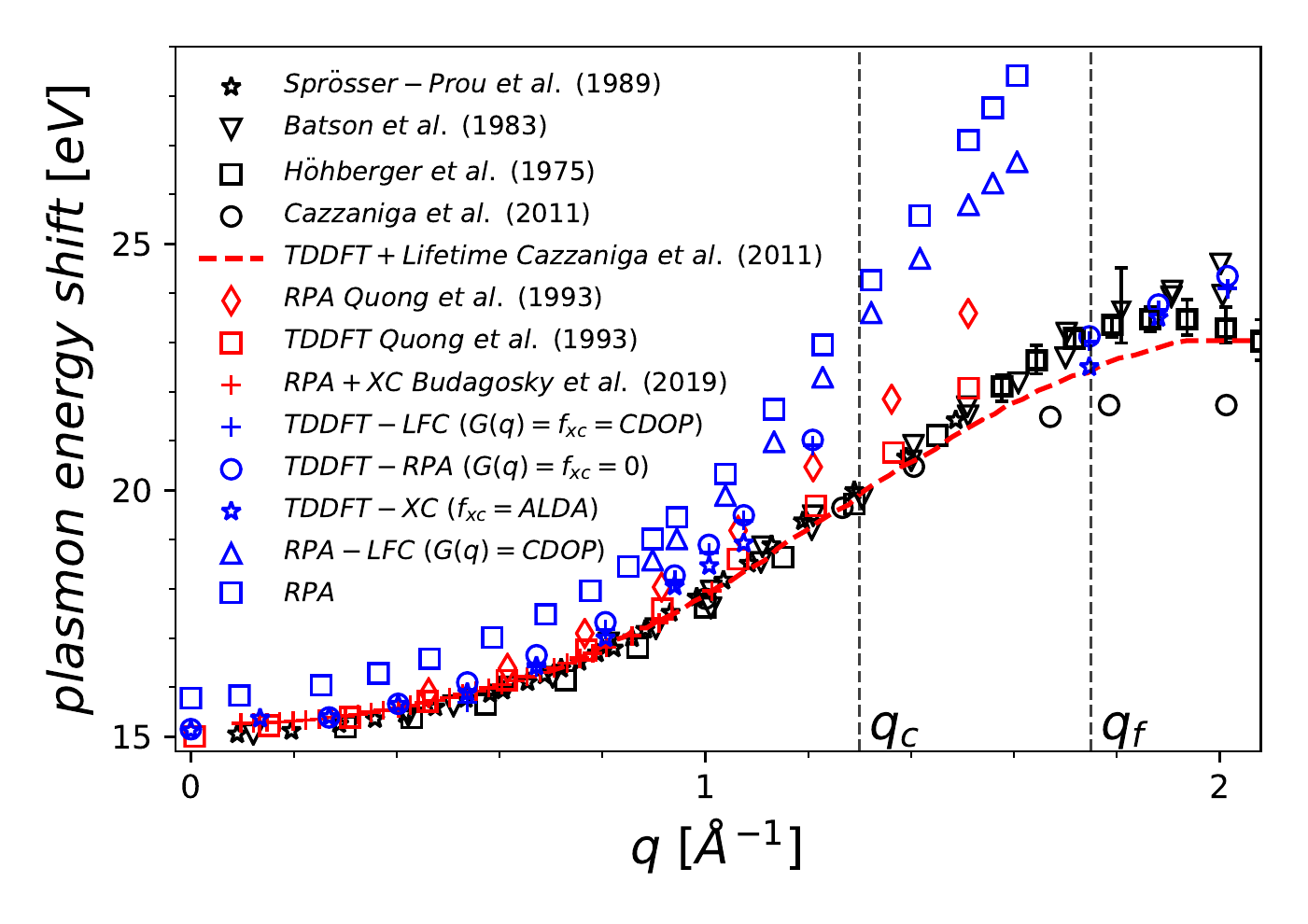}   
\caption{\raggedright Aluminum plasmon dispersion under ambient conditions. The critical and the Fermi wavenumbers are indicated by the vertical lines. Experimental data shown in black symbols stems from Ref.~\onlinecite{PhysRevB.40.5799, PhysRevB.27.5224, PhysRevB.84.075109, hohberger1975plasmon}. Theoretical data shown in red symbols are taken from Ref.~\onlinecite{PhysRevB.84.075109, PhysRevB.99.245149, PhysRevB.49.2362, PhysRevLett.70.3955}. TDDFT-RPA/TDDFT-XC/TDDFT-LFC/RPA-LFC/RPA results of this work are indicated with blue symbols.  }    
\label{Al_dispersion}     
\end{figure} 

The plasmon dispersion under ambient conditions is shown in Fig. \ref{Al_dispersion}. The plasmon disperses quadratically up to the critical wavenumber ($q_{c}$) and a flattening is observed for $q>q_{c}$. $q_{c}$ is defined as the wavenumber at which the dispersion merges into the continuum of the single-particle excitations~\cite{pines2018theory,ichimaru1982strongly}. For very small wavenumbers, in the optical limit, the Landau damping is very small~\cite{landau1946vibrations} and the decay of the plasmon is mainly due to band structure effects ~\cite{ichimaru1982strongly}. Electron-electron interactions play a stronger role with increasing wavenumber~\cite{hamann2020ab}. For wavenumbers above $q_{c}$, a plasmon cannot be defined based on many-particle dielectric theories~\cite{hamann2020ab}, hence a shift based on the location of the peak of the DSF is given (Fig.~\ref{Al_dsf_ambient_conditions}). 

For small \emph{q}, the various TDDFT approaches agree well with the theoretical results of Quong \textit{et al.}~\cite{PhysRevLett.70.3955} and the experimental measurements by Spr\"osser-Prou and Batson \textit{et al.}~\cite{PhysRevB.40.5799, PhysRevB.27.5224}. Near $q_c$, we start to see deviations between experiment and different theoretical results, as already mentioned in the discussion of Fig.~\ref{Al_dsf_ambient_conditions}. This is due to the broadening of the plasmon peak and the onset of two-peak features in the DSF, which complicate determining the peak position without determining the zeros of the dielectric function~\cite{hamann2020ab}. Therefore, we do not provide TDDFT results for the shift in the intermediate range.

For large \emph{q}, the experimental results obtained by Batson~\cite{PhysRevB.27.5224} and H\"ohberger~\cite{hohberger1975plasmon} \textit{et al.} agree with our results. In this case, one should not speak of a plasmon anymore. The observed feature is better described by a shift of the peak of the DSF that is now dominated by single-particle excitations.
The results of Batson~\cite{PhysRevB.27.5224} \textit{et al.} show a flattening in the plasmon dispersion curve only for larger $q$ as plotted.

The inclusion of different LFCs (TDDFT-LFC, CDOP Corradini \textit{et al.}~\cite{PhysRevB.57.14569}), and XC Kernels (TDDFT-XC) results in a lowering of the plasmon shift at intermediate and large \emph{q}. The influence of the TDDFT XC kernel compared to RPA in the lowering of the plasmon shift is also observed in the theoretical results of Quong and Cazzaniga \textit{et al.}~\cite{PhysRevLett.70.3955,PhysRevB.84.075109}. Further improvements to the ALDA kernel can be achieved by considering an exact-exchange kernel (EXX)~\cite{marques2006time}. The inclusion of lifetime effects in TDDFT lowers the shift further as shown by Cazzaniga \textit{et al.}~\cite{PhysRevB.84.075109}. However, the experimental results at large wavenumbers by Cazzaniga {\em et al.}~\cite{PhysRevB.84.075109} seem to contradict the results of Sprösser~\cite{PhysRevB.40.5799}, Batson~\cite{PhysRevB.27.5224}, and Höhberger~\cite{hohberger1975plasmon}. This mainly illustrates the difficulty of extracting peak positions from structure factors at large \emph{q}. 

\subsubsection{Plasmon Lifetimes}  
\begin{figure}[H]          
\centering     
\includegraphics[width=1.0\columnwidth]{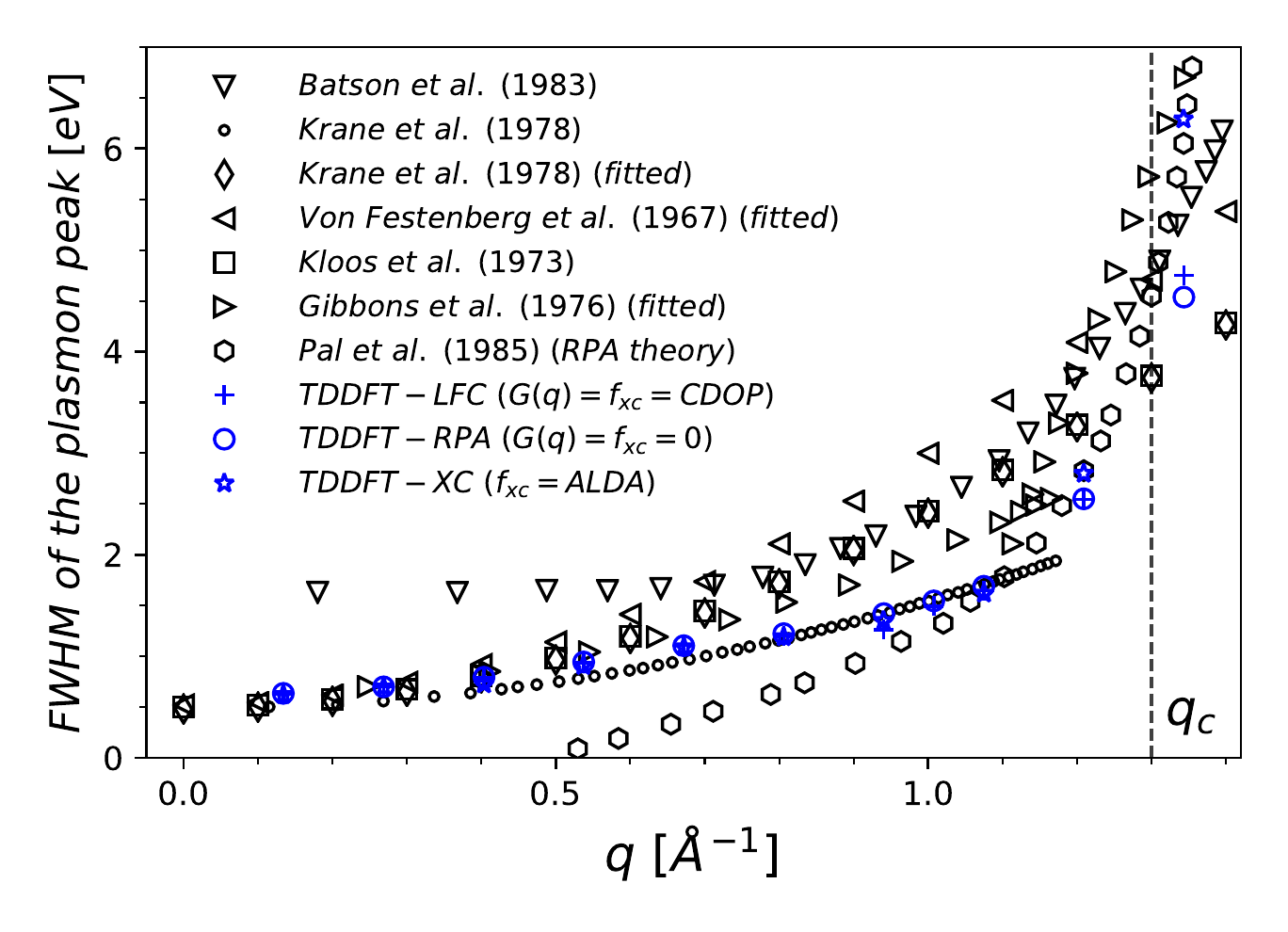} \caption{\raggedright Aluminum plasmon peak FWHM under ambient conditions. The critical wavenumber is indicated by the vertical line. Experimental and theoretical data shown in black symbols stems from Ref.~\onlinecite{Krane_1978, PhysRevB.27.5224, Festenberg1967, Kloos1973, gibbons1976line, pal1985plasmon}. TDDFT-RPA/TDDFT-XC/TDDFT-LFC results of this work are indicated with blue symbols. }   
\label{Al_width}
\end{figure}

The full width at half maximum (FWHM) of the plasmon is shown in Fig. \ref{Al_width}. This quantity reflects the strength of plasmon damping as can be extracted from the Lorentz profile of the weakly damped plasmon at small wavenumbers. It can also be determined by finding the zeros of the complex dielectric function $\epsilon[q,\omega(q)-i\gamma(q)]$~\cite{ichimaru1982strongly,hamann2020ab} in which the imaginary parts correspond to inverse lifetimes.

As the FWHMs computed using LR-TDDFT depend on $\eta$, it is necessary to consider how to extract this quantity consistently. We extrapolate the value of the FWHM for multiple values of $\eta$ to the $\eta \rightarrow 0$ limit for a sufficiently dense sampling of both the energy domain and the first Brillouin zone~\cite{PhysRevB.96.165417}.
 
The data is shown up to the wavenumbers near $q_{c}$ where a stable plasmon feature is obtained from $S(q,\omega)$. The width calculated within TDDFT-RPA and TDDFT-XC has a flat feature for $q<1.0/ \si{\angstrom}$ and then grows rapidly with increasing $\emph{q}$ which can also be seen in the experimental measurements~\cite{Krane_1978,Kloos1973,PhysRevB.27.5224,Festenberg1967}. The inclusion of LFC increases the width for \emph{q} above $q_{c}$ and has negligible impact for small \emph{q} where the width is dominated by the band structure as calculated in the DFT calculations. This is exemplified by the good agreement of TDDFT-RPA and TDDFT-XC for $q<1.0/ \si{\angstrom}$. Significant deviations between the two emerge near $q_{c}$ when the LFC has an increasing impact. However, the deviations between TDDFT-RPA and TDDFT-XC for the plasmon dispersion starts to appear at much smaller wavenumbers. 
 
While our calculated plasmon dispersion curves are in good agreement with the results of Batson \textit{et al.}~\cite{PhysRevB.27.5224}, the lifetimes/decay rates given by Batson deviate from our results. Our plasmon lifetime results are in best agreement with the experimental results of Krane \textit{et al.}~\cite{Krane_1978} and, at small \emph{q}, with the experimental results of Kloos \textit{et al.} and Von Festenberg \textit{et al.}~\cite{Kloos1973,Festenberg1967}. 

Any experimental measurement, e.g., via XRTS, gives a \emph{q}-dependent scattering signal featuring a plasmon energy shift and a width associated with it. Information on both of these parameters are vital to benchmark (dynamic) LFCs, collision frequencies, and kernels in order to produce accurate TDDFT models. However, most experimental results available to us for ambient aluminum provide either the dispersion or the decay rates with the exception of Batson \textit{et al.}~\cite{PhysRevB.27.5224}. Thus, with the Batson data in its entirety not being consistent with our results and the lack of further consistent plasmon position and FWHM data from experiments, it yields an inconclusive, hence, very unsatisfactory picture. We are not able to compare both plasmon position and FWHM of the plasmon peak to other theoretical predictions either due to a lack of data.   

\subsection{Extreme Conditions}
Measurements of the plasmon dispersion and plasmon peak FWHM at extreme conditions of high pressure and temperature are quite challenging. Using isochoric heating by optical or X-ray pulses, solid aluminum foils can be heated to high temperatures. Combining such a setup with X-ray and optical diagnostics, the electronic response of WDM can be accessed~\cite{PhysRevLett.115.115001}. Higher densities and, therefore, higher pressures can also be reached via isentropic or shock compression using high intensity laser pulses~\cite{Fletcher2014,doi:10.1063/1.5070140}. 
The technical details of our (TDDFT) calculations for WDM conditions are listed in Appendix~\ref{app_Dl}.  

\begin{figure}[htp]                  
\centering             
\includegraphics[width=1.0\columnwidth]{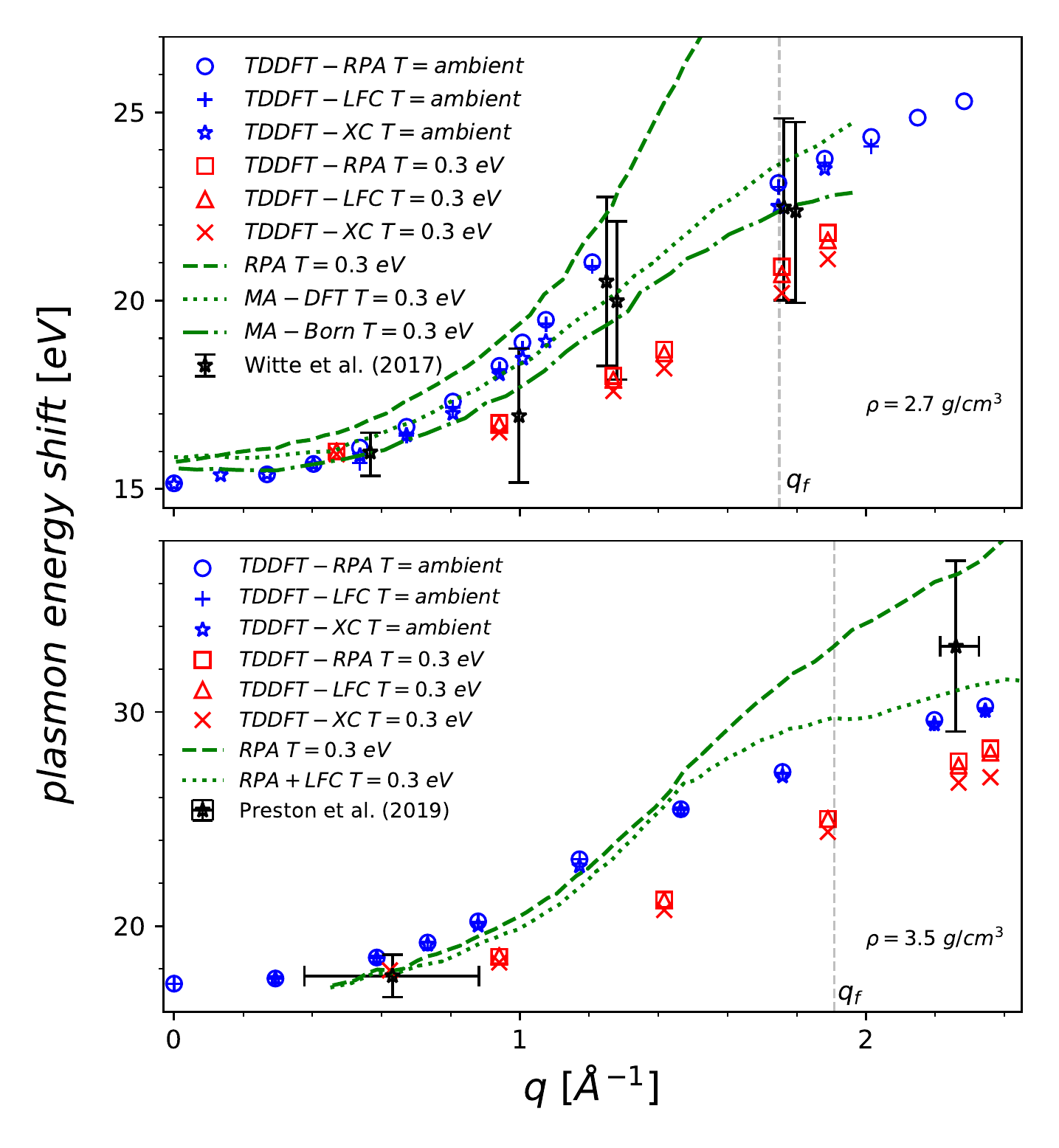} 
\caption{\raggedright Aluminum plasmon dispersion under extreme conditions (T=0.3 eV) for densities of 2.7 g/$cm^{3}$ (uncompressed) and 3.5 g/$cm^{3}$ (compressed). The Fermi wavenumber is indicated by the vertical line. Experimental and theoretical data for RPA, MA, RPA+LFC (in green) stems from Ref.~\onlinecite{PhysRevLett.118.225001,doi:10.1063/1.5070140}. TDDFT-RPA, TDDFT-XC, and TDDFT-LFC results of this work are indicated with blue and red symbols.} 
\label{Al_dispersion_extreme}
\end{figure}

\subsubsection{Plasmon Dispersion}
In Fig.~\ref{Al_dispersion_extreme}, the plasmon dispersion is shown for densities 2.7 g/cm$^{3}$ (uncompressed) and 3.5 g/cm$^{3}$ (compressed) at ambient temperature and at $T=0.3$ eV. Under these conditions, the temperature should have negligible impact on the plasmon dispersion, because it depends primarily on the electron density when the Fermi energy exceeds the temperature (i.e., for small $\Theta$). However, the quadratic term in Eq.~(\ref{plasmon_dispersion}) is generally temperature dependent. The influence of any finite-temperature LFC (T-LFC), $G(q, r_{s}, \Theta)$, can be readily assessed based on the density, temperature and the momentum vector of the system (Table \ref{t: Al_lfc_table} in Appendix \ref{app_lfc}). Due to extremely small $\Theta=0.02-0.025$, temperature effects can be ignored in $G(q,r_{s},\Theta) \rightarrow G(q,r_{s})$. To this end, we also perform a comparison with RPA and TDDFT results computed at ambient temperature. 

When the static LFC is included, the plasmon dispersion is reduced at large \emph{q} and approaches the results obtained with TDDFT akin to the LFC approximation used for the XC kernel in TDDFT. Within the RPA, we also investigated the effect of treating the electrons within an all-electron formalism rather than a pseudopotential-based formalism. We found that including the core electrons on a system size up to $N=32$ yields only an insignificant deviation on the shift from those calculated with the use of a pseudopotential.  

We compare our data for the uncompressed case at nominal $T=0.3$~eV with both the experimental measurements (black symbols) and the theoretical plasma physics models (green curves) of Witte \textit{et al.}~\cite{PhysRevLett.118.225001}. For small wavenumbers, all our TDDFT results agree well with the experimental and other curves, which is mainly an indication that the density is correct. At larger wavenumbers, deviations are apparent which are caused by differing temperatures and different levels of approximations. Due to the large error bars, it is not possible to outright discard any theory with the exception of the pure RPA (green dashed). However, it seems that within the TDDFT results, there is no indication of the temperature being as extracted by Witte {\em et al.}~\cite{PhysRevLett.118.225001}. The $T=0.3$~eV results (red) seem consistently on the lower end of error bars of the measurements. A better agreement is reached when considering the ions at ideal lattice positions and not in a molten state (blue symbols). This seems reasonable, as the time frame of the measurements is in the $100$~fs range and, hence, too short for the onset of any significant ion motion. This demonstrates the problem of the model-dependent temperature extraction in such experiments~\cite{PhysRevLett.120.205002,PhysRevB.102.195127}. 

\begin{figure}[htp]             
\centering              
\includegraphics[width=1.0\columnwidth]{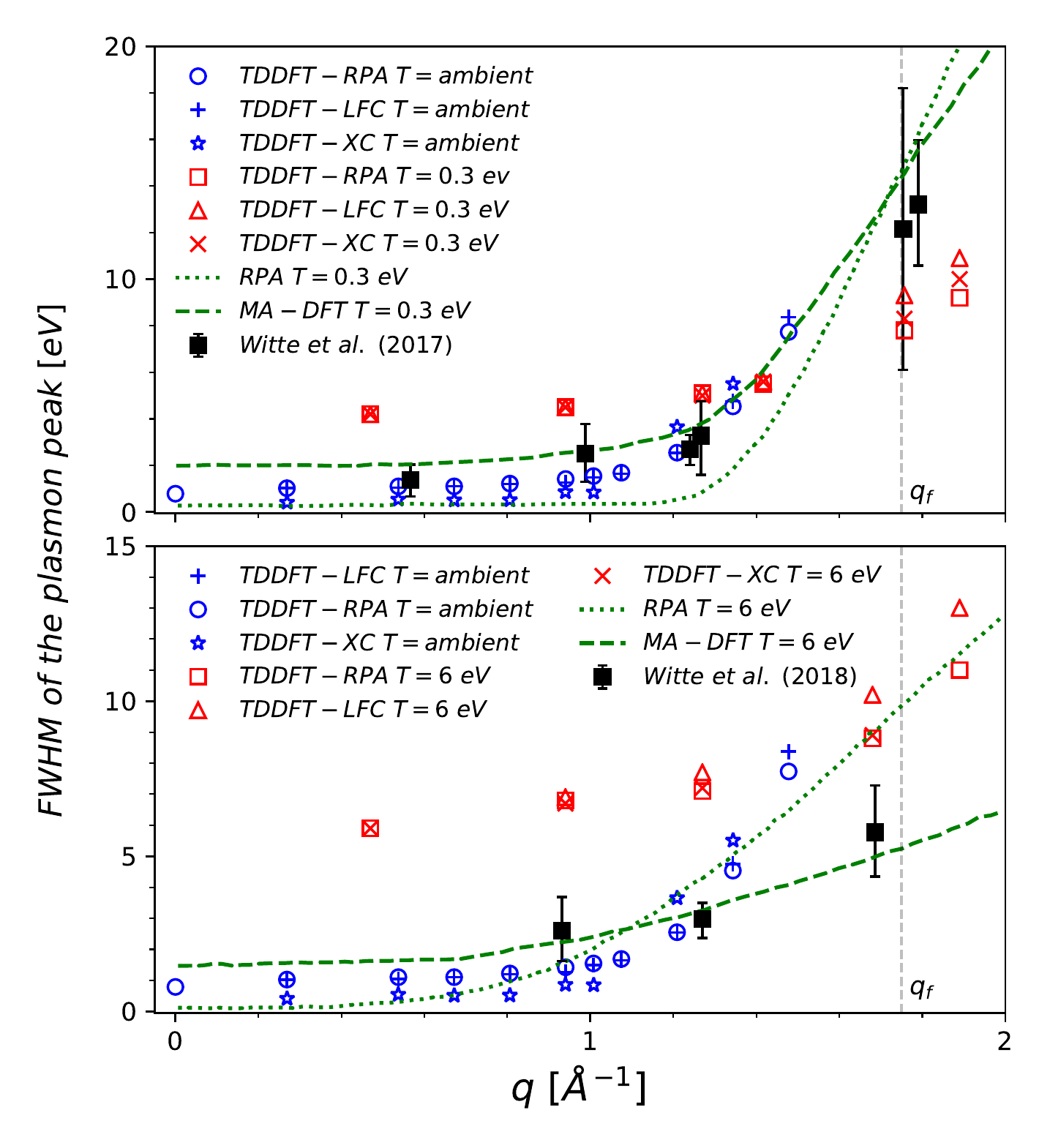}  \caption{\raggedright Aluminum plasmon peak FWHM under extreme conditions for $\rho=2.7 \ g/cm^{3}$ at $T=0.3$ eV (above) and $T=6$ eV (below). The Fermi wavenumber is indicated by the vertical line. Experimental and theoretical data for RPA, MA (in green) stems from Ref.~\onlinecite{PhysRevLett.118.225001, doi:10.1063/1.5017889}. TDDFT-RPA, TDDFT-XC, and TDDFT-LFC results of this work are indicated with blue and red symbols.} 
\label{Al_width_extreme}  
\end{figure}

In the compressed case, the available data set is restricted to two measurements, i.e., at small and large \emph{q}~\cite{doi:10.1063/1.5070140}. At small \emph{q}, the data agrees well with the experimental measurement which in this case is not trivial due to the shocked state of the system. Thus, the density determination seems reasonable. At large \emph{q}, the TDDFT results are much lower than the experimental results due to the strong influence of the ions, which are in a liquid or plasma like state, at the elevated temperature. Ignoring the temperature of the ions (ideal lattice at ambient temperatures), the simulations are in better agreement with the experimental plasmon shift. 

The transition from a quadratic dispersion to a flat feature can be observed at a smaller \emph{q} when compared to aluminum at ambient density.
RPA+LFC theory and TDDFT results indicate an increased damping with a raise in density (top versus lower panels in Fig.~\ref{Al_dispersion_extreme}), but the experimental data remains inconclusive.

In summary, we find that the ambient data is in much better agreement with both XRTS measurements than the results obtained at $T=0.3$ eV~\cite{PhysRevLett.120.205002}. Of course, temperature measurements via XRTS, if not done via detailed balance, are always model dependent. We stress that LR-TDDFT using appropriate XC kernels or LFCs, respectively, is far more capable of including electron-electron as well as electron-ion correlations in the computation of collective effects and structure factors than any other theory.


\begin{figure}[H]
\centering   
\includegraphics[width=1.0\columnwidth]{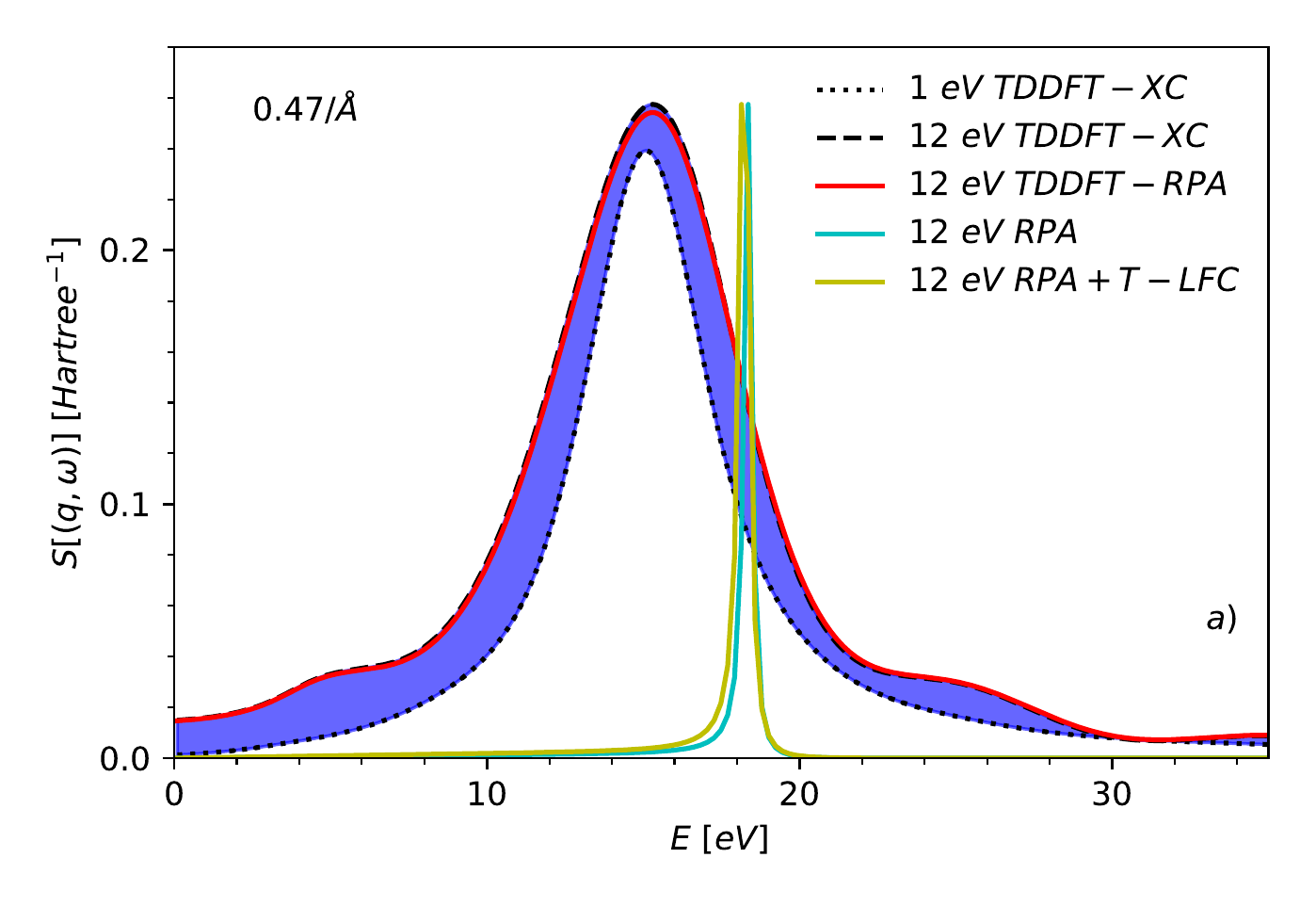}
\includegraphics[width=1.0\columnwidth]{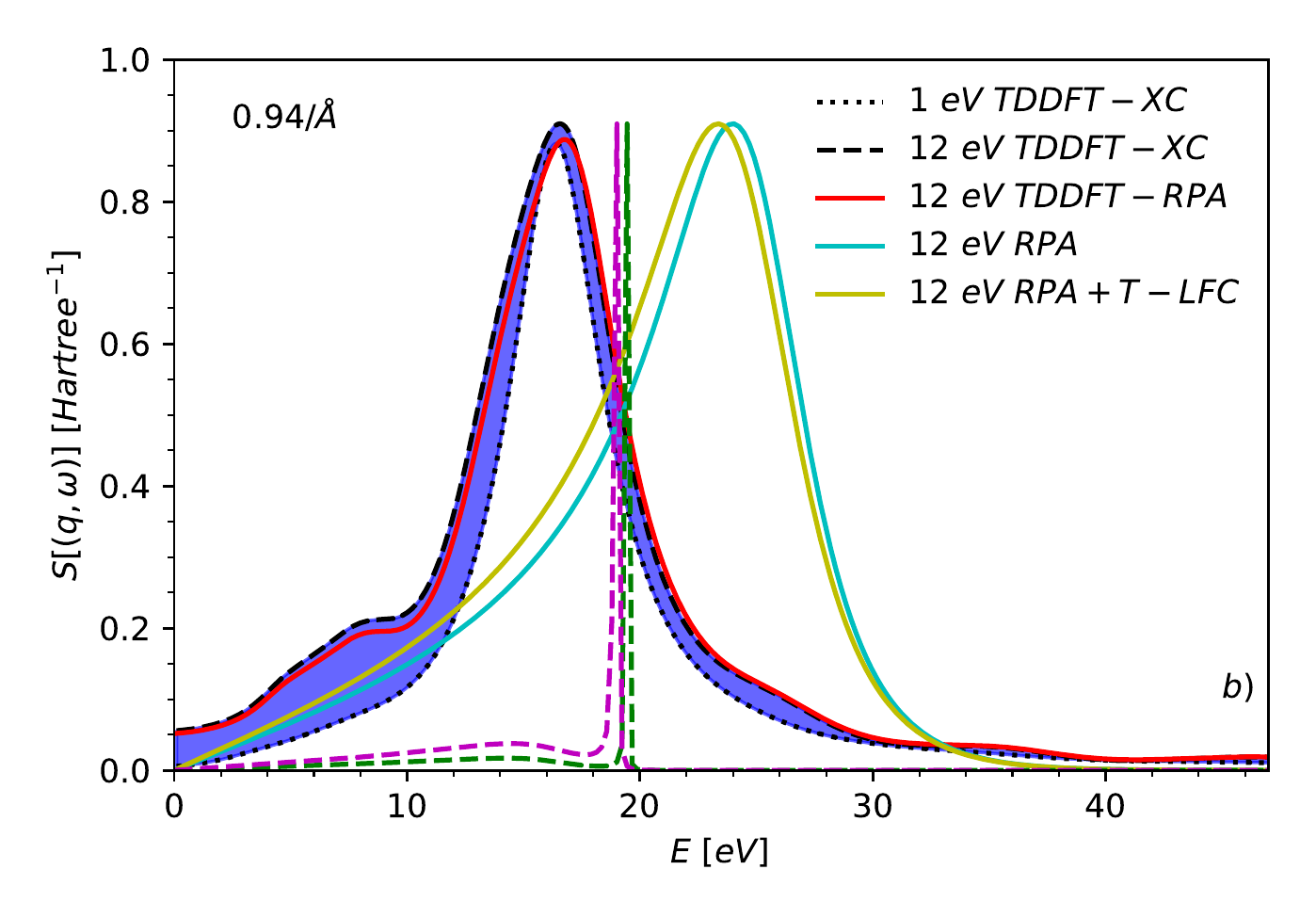}
\includegraphics[width=1.0\columnwidth]{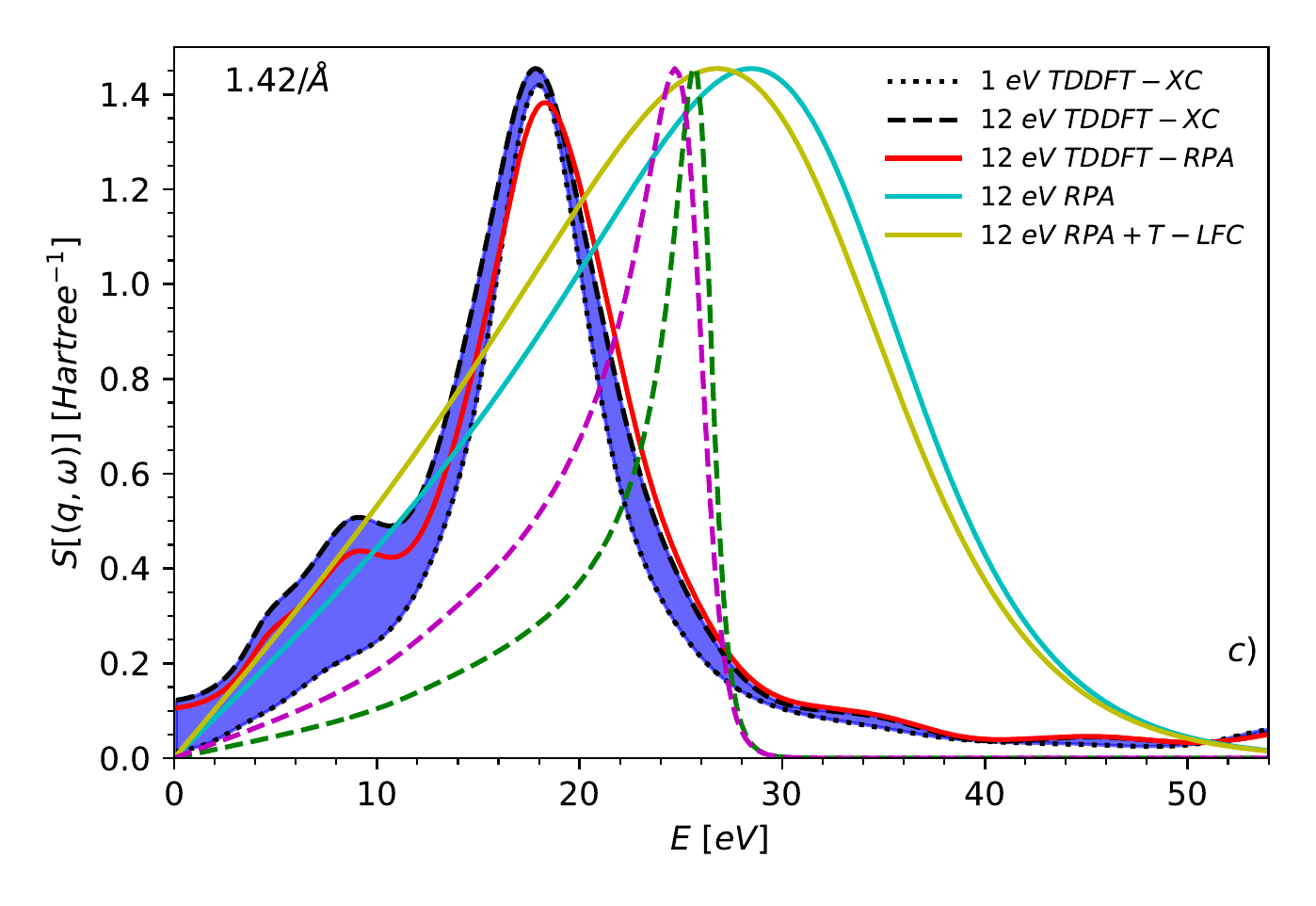}
\caption{\raggedright DSF for aluminum in atomic units for a) 0.47, b) 0.94, and c) 1.42 (\si{\angstrom^{-1}}) at various temperatures. The TDDFT-RPA results are shown only at 12 eV. The TDDFT-XC results are shown from 1 to 12 eV in the blue area (black curves for 1 and 12 eV) and are broadening with temperature. The purple and green dashed lines are the RPA and RPA$+$T-LFC results at 1 eV respectively. The RPA results are normalized with respect to the TDDFT-XC results at 12 eV to allow plotting them at the same scale. } 
\label{Al_dsf_temp}  
\end{figure}       


\subsubsection{Plasmon Lifetimes}

In Fig.~\ref{Al_width_extreme}, the plasmon FWHM is shown for a density of 2.7 g/cm$^{3}$ and temperatures of $0.3$ and $6.0$ eV. The data is compared to the available experimental results of Witte \textit{et al.} and their calculations using plasma theory~\cite{PhysRevLett.118.225001, doi:10.1063/1.5017889}. Nominal at $T=0.3$ eV, the experimental data agrees well with the TDDFT results for the cold case. The width resulting from TDDFT in the cold case features a similar trend than the free electron gas where both methods agree for small \emph{q}. At $T=0.3$ eV, the width is obtained from the linear response calculations involving DFT-MD snapshots. Here, the TDDFT results feature larger widths for small \emph{q} but the data still lies within the large error bars of the experimental results at large \emph{q}. 

A similar trend can be observed for the case of $T=6$~eV as presented in the bottom panel of Fig.~\ref{Al_width_extreme}. The cold TDDFT results fit the experimentally determined width much better than the TDDFT data at elevated temperatures where the width is increased strongly due to the liquid structure of the ions.

Remarkably once again, the cold data is in much better agreement with the XRTS measurements than results at $0.3$~eV and $6$~eV data. Apart from the model-dependent temperature determination as mentioned above, this hints at the fact that the experimental time scales are too short to allow an equilibrium of the coupled electron-ion system to be established.      

\subsubsection{Temperature Dependence of the DSF}
In Fig.~\ref{Al_dsf_temp}, the DSF of aluminum for various temperatures is shown. The usual dispersion and change in lifetime of the plasmon can be observed in panels a) to c). For TDDFT, we notice that as long as the plasmon dominates the spectrum [panels a) and b)], the locations of the peaks at a specific \emph{q} are independent of temperature. This is contrary to the prediction of the Lindhard-RPA (with and without LFCs, cyan and olive curves, respectively), for which the energy of the plasmon changes drastically with temperature in panel b). Of course, temperature results in an increase of width of the plasmon peak. Once single-particle effects start to influence the structure factor for larger wavenumbers as in panel c), temperature causes a change in the position of the peak as predicted by TDDFT, too. 

The inclusion of LFCs yields a downshift of the intensities to lower frequencies. Within TDDFT, also a slight increase of the peak height at large wavenumbers shown in panel c) is observed. 
 
The influence of the finite-temperature LFC is only apparent at large \emph{q} and at high temperatures, that is at 12~eV where a deviation from ground state LFC is observed in the energy range 0-10~eV. Table. \ref{t: Al_lfc_table} in Appendix~\ref{app_lfc} summarizes the LFCs considered in this work.     
We conclude that the temperature determination from the plasmon peak and width is highly model dependent and great care should be taken in the choice of the applied theory.

\subsubsection{Static Structure Factor}  
An important test of the quality of the DSF as presented in the preceding sections is given by the calculation of several different moments of the structure factor. Here, we focus on the calculation of the static structure factor from the TDDFT spectra according to
\begin{equation}
S(q) = \int\displaylimits_{- \infty }^{ \infty } S(q,\omega) d\omega .
\end{equation}

\begin{figure}[htp]             
\centering      
\includegraphics[width=1.0\columnwidth]{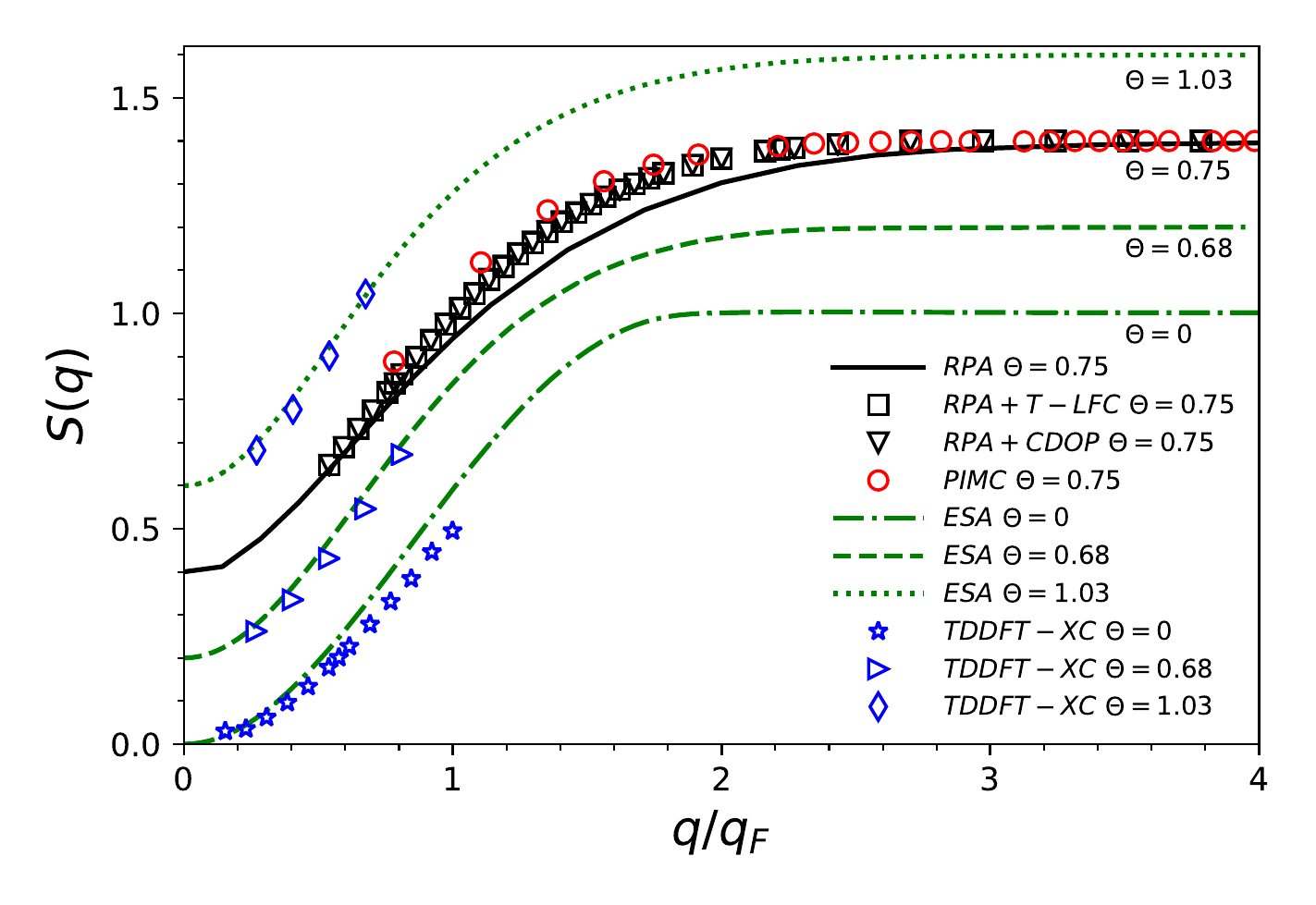} \caption{\raggedright Static structure factor for aluminum ($r_{s}=2.07$) computed using PIMC and RPA  including LFCs at $\Theta=0.75$ ($T=8.77$ eV). The ESA results are shown for $\Theta=0$, $\Theta=0.68$ ($T=8$ eV) and $\Theta=1.03$ ($T=12$ eV). The TDDFT-XC results are shown from ambient to $T=12 \ eV$ for $q/q_{F}$ up to $\sim 1.0$.  } 
\label{Al_ssf}          
\end{figure}

In Fig. \ref{Al_ssf}, the electronic static structure factor of aluminum ($r_{s}=2.07$) within different theories is shown for several values of the degeneracy temperature $\Theta$. The red circles correspond to PIMC results for the uniform electron gas at $\Theta=0.75$ and are compared to RPA calculations including static LFCs both at finite temperature (T-LFC) and in the ground state (CDOP). The TDDFT-XC results are also shown for comparison in the temperature range up to 12 eV for $q/q_{F} \lesssim 1.0$. Note that only the contribution of the valence electrons is considered in the TDDFT-XC calculations. Furthermore, the TDDFT-XC results are limited to the displayed range of wavenumbers, because at higher values there are other excitations (\emph{L}-edge, specifically with $L_{2,3}$ and $L_{1}$) that do not occur in an electron gas as considered in PIMC. Finally, the green curves have been obtained using the novel effective static approximation (ESA)~\cite{dornheim2020effective}, which has been shown to yield highly accurate results for $S(q)$ over the entire WDM regime, with a typical systematic error of $\sim0.1\%$ as compared to PIMC.


The agreement between the integrated TDDFT spectra and the static structure from both PIMC and ESA is very satisfactory and can serve as a benchmark of the quality of the TDDFT spectra. However, it should also be noted that integrated quantities like the static structure factor are prone to hiding inaccuracies of the dynamic quantities~\cite{dornheim2020effective}.

\section{Conclusions} 
We demonstrated the capabilities of LR-TDDFT in calculating plasmon dispersion and plasmon lifetimes for the simple metal aluminum at ambient and extreme conditions. We studied aluminum as a perfect fcc lattice as well as a high pressure fluid. We used both all-electron codes and PAW pseudopotentials. Starting from TDDFT-RPA, we used a variety of XC kernels in the LR-TDDFT equations: ALDA, static $T=0$ LFCs, and temperature-dependent LFCs, the latter two based on QMC simulations.

We compared our results to other TDDFT results and to plasma physics theories using the Mermin dielectric function and several different collision frequencies. Also, where available, we compared to experimental values.

Our analysis is based on relatively few complete data sets of both plasmon lifetimes and plasmon dispersion (of which we present one) for aluminum at room temperature. Within this dataset, there is basically no consistent case in which two theories (or experiments) agree in plasmon location and lifetime simultaneously. It is even more worrisome that TDDFT calculations that should be capable of obtaining very similar results (based on the published set of parameters and methods) fail to do so. While this is the case for aluminum at ambient conditions, the situation is naturally worse for warm dense, or high temperature aluminum where the error bars and uncertainties are larger due to experimental difficulties and computational challenges.

This has significant repercussions for the evaluation of experimental spectra from XRTS and other experiments, because such spectra are also used for temperature and density determination of the created states. While this is less problematic for states under ambient conditions or at high pressure in solids, it is a challenge for WDM states. XRTS is, in principle, one of the very few methods capable of obtaining such basic parameters which are used as input to subsequent simulation techniques. We, therefore, not only need accurate and reliable methods to calculate the dynamic structure but also fast methods to be able to fit spectra. Our assessment clearly points to a strong need for the development or improvements in reliable methods such as in LR-TDDFT.

\begin{acknowledgments} 
We are grateful to Maximilian B\"{o}hme for fruitful discussions and helpful comments. We thank Thomas Preston for the correspondence on experimental details regarding compressed aluminum. We are also grateful to Chongjie Mo for providing the simulation data on aluminum. KR, TD and AC acknowledge funding by the Center for Advanced Systems Understanding (CASUS) which is financed by the German Federal Ministry of Education and Research (BMBF) and by the Saxon Ministry for Science, Culture and Tourism (SMWK) with tax funds on the basis of the budget approved by the Saxon State Parliament. Computations were performed on a Bull Cluster at the Center for Information Services and High Performance Computing (ZIH) at TU Dresden. We would like to thank the ZIH for its support and generous allocations of compute time. Sandia National Laboratories is a multimission laboratory managed and operated by National Technology \& Engineering Solutions of Sandia, LLC, a wholly-owned subsidiary of Honeywell International Inc., for the U.S. Department of Energy’s National Nuclear Security Administration under contract DE-NA0003525.

\end{acknowledgments}    

\clearpage

\section*{Appendix}

\appendix

\section{Computational Details}
\label{app_Dl}

\begin{figure}[htp]                     
\centering             
\includegraphics[width=1.0\columnwidth]{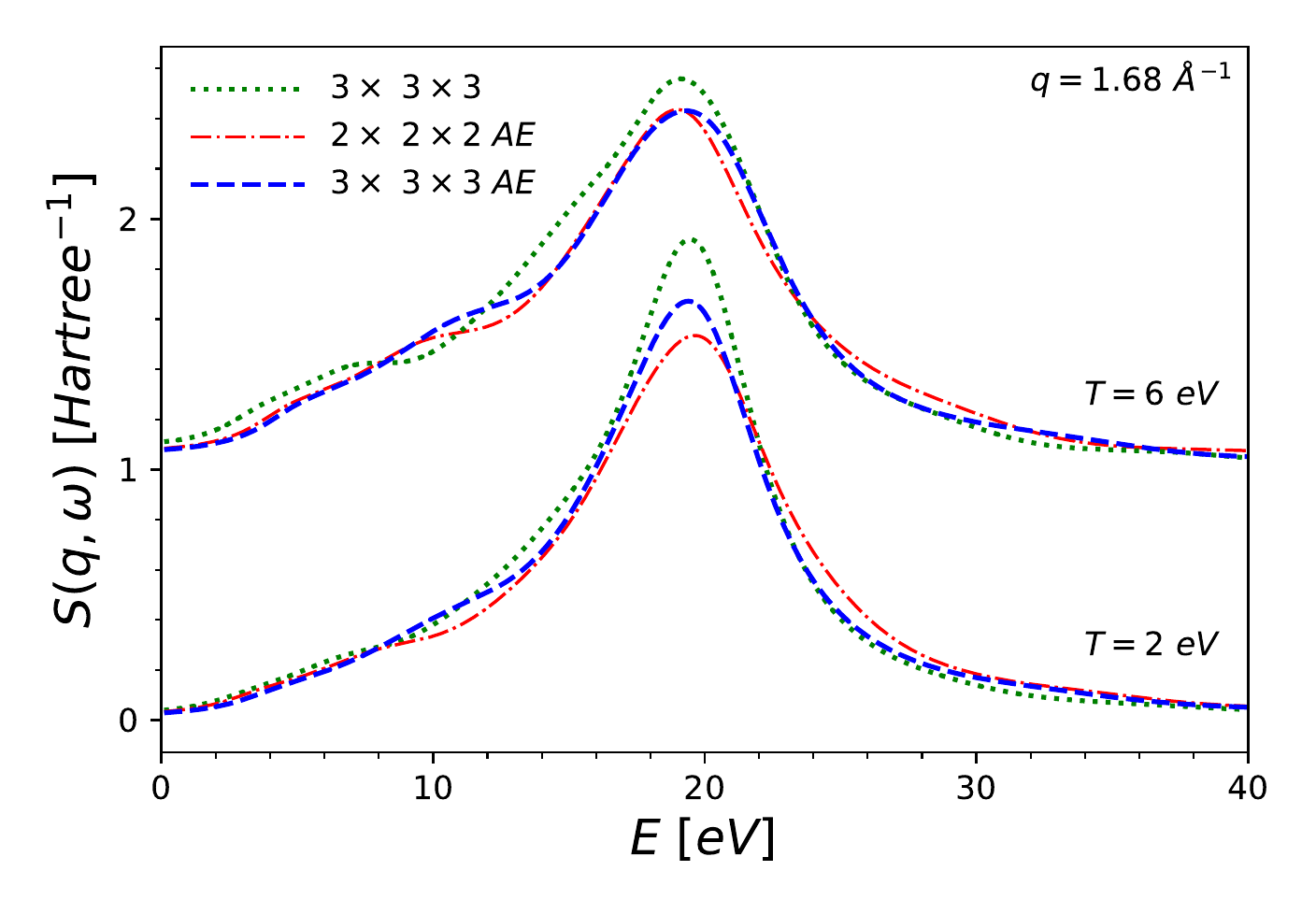}      
\caption{\raggedright DSF for aluminum ($\rho=2.7 \ g/cm^{3}$) in atomic units at $T=2$ eV and $T=6$ eV  with respect to \emph{k}-points and the number of electrons considered in the pseudopotential. AE refers to the use of an all electron pseudopotential (11 electrons ignoring the $1s^{2}$ core) compared to the 3 valence electrons otherwise.}    
\label{Al_2ev_lanczos}  
\end{figure}  

At ambient conditions (T = 300 K), the LR-TDDFT calculations were performed using the full-potential linearised augmented-plane wave code implemented in elk~\cite{elk}. A \emph{k}-point grid of $40\times 40\times 40$ points was used and $80$ bands were considered for the fcc unit cell. Fermi smearing was used with a width of $0.01$~Ha. The adiabatic local density approximation (ALDA) as implemented in the elk code was used.

At WDM conditions, DFT-MD simulations were performed using VASP~\cite{PhysRevB.47.558,PhysRevB.59.1758,KRESSE199615,PhysRevB.54.11169}.
We used PAW pseudopotentials~\cite{blochl1994projector} with three electrons considered valence and a core radius of $r_c=1.7$~a$_B$. The plane wave cutoff was set to $350$~eV and the convergence in each self-consistency cycle was set to $10^{-5}$. We used the Mermin formulation of thermal DFT~\cite{Mermin_1965} and Fermi occupation of the eigenvalues. Generally, the first Brillouin zone was sampled on a $2\times 2\times 2$ grid of \emph{k}-points. The number of bands varied with the temperature up to $1050$ for the highest temperature of $T=12$~eV in a $N=32$ supercell. Both LDA~\cite{KoSh1965} and PBE~\cite{perdew1996generalized} XC functional were used. The thermostat in the NVT ensemble was of Nose-Hoover type~\cite{PhysRevA.31.1695}. Ionic time steps of $\Delta t = 0.2$~fs were taken. 

Simulations involving large system sizes at high temperatures and pressures are computationally too expensive within a full-potential linearised augmented-plane wave code. The KS orbitals for a supercell containing 32 aluminum atoms were, therefore, generated from DFT calculations on pseudopotential within Quantum ESPRESSO electronic structure code~\cite{ giannozzi2009quantum,giannozzi2017advanced}. The LDA norm-conserving pseuopotentials were generated with the OPIUM package~\cite{opium}. 11 valence electrons were considered in the psuedopotential, while the $1s^2$ core is ignored. The plane-wave cutoff to represent the KS orbitals is set to 70 Ry. Electronic occupations are generated using a Methfessel-Paxton smearing~\cite{PhysRevB.40.3616} where the number of bands at a temperature of 12 eV is set to roughly 1050. The Brillouin zone was sampled using $ 3 \times 3 \times 3 $ Monkhorst-Pack mesh throughout. Based on these KS orbitals as input, LR-TDDFT were performed with the yambo~\cite{MARINI20091392}, turboTDDFT~\cite{turbotddft} and TDDFPT~\cite{PhysRevB.88.064301,TIMROV2015460} packages.

The static limit of the LFCs is substituted in Eq.~(\ref{eq:define_LFC}) as $G(q,r_{s})$ at ground state and as $G(q,r_{s},\Theta)$ at finite temperature.

\section{Convergence Analysis of the DSF} 
\label{app_Al}
 
The convergence with respect to the number of \emph{k}-points and bands is important due to the computational cost. The DSF for aluminum at 2 eV and 6 eV using 32 atoms for 600  and 750 bands respectively is shown with respect to the \emph{k}-points and the number of electrons considered in the pseudopotential in Fig. \ref{Al_2ev_lanczos}. The calculations are well converged with respect to the number of \emph{k}-points. The all-electron (AE) LDA Perdew-Zunger norm-conserving pseudopotential results in a lowering of the peak intensity and an increase in the intensity to higher energies at the shoulder for $T=6$ eV at frequencies near 10 eV.

\section{Details on the Local Field Corrections} 
\label{app_lfc} 

The LFCs for aluminum at ambient and compressed densities (2.7 and 3.5 g/$cm^{3}$) in this work is shown in Table. \ref{t: Al_lfc_table}.

\begin{table}[htp]  
\centering
\begin{tabular}{lcccc} 
\hline
\hline    
$\rho \ (g/cm^{3})$ & $T \ (eV)$  & $q \ (\si{\angstrom^{-1}}) $  & LFC & T-LFC  \\
\hline   
\hline  
2.7 & 1.0 & 3.02 &  0.79  & 0.79  \\
2.7 & 3.0 & 3.02 &  0.79  & 0.79  \\
2.7 & 6.0 & 3.02 &  0.79  & 0.76  \\
2.7 & 8.0 & 1.89 &  0.33  & 0.34  \\
2.7 & 8.0 & 2.36 &  0.53 & 0.51  \\
2.7 & 8.0 & 2.83 &  0.72 & 0.67  \\ 
2.7 & 8.0 & 3.02 &  0.79 & 0.74  \\ 
2.7 & 12.0 & 0.47 &  0.02 & 0.02  \\
2.7 & 12.0 & 0.94 &  0.08 & 0.09  \\
2.7 & 12.0 & 1.42 &  0.19 & 0.20  \\
2.7 & 12.0 & 1.89 &  0.33 & 0.34  \\
2.7 & 12.0 & 2.36 &  0.53 & 0.48  \\
2.7 & 12.0 & 2.83 &  0.72 & 0.63  \\
3.5 & 0.3 & 1.89 &  0.29 & 0.29  \\ 
3.5 & 0.3 & 2.36 &  0.45 & 0.45  \\ 
3.5 & 0.3 & 2.83 &  0.63 & 0.63  \\ 
3.5 & 0.3 & 3.02 &  0.70 & 0.70  \\ 
\hline     
\hline 
\end{tabular}   
\caption{\raggedright Local field corrections (LFC) and finite-temperature LFCs (T-LFC) for aluminum at 2.7 and 3.5 g/$cm^{3}$ for various temperatures and wavenumbers.} 
\label{t: Al_lfc_table} 
\end{table}      

\bibliography{plasmon-bibliography}
 
\end{document}